\documentclass[preprint, 12pt]{elsarticle}

\usepackage{amsmath}
\usepackage{amssymb}
\usepackage{pifont}
\usepackage[colorinlistoftodos]{todonotes}
\usepackage{natbib}
\biboptions{authoryear}
\usepackage{geometry}
\usepackage{fleqn}
\usepackage{graphicx}
\usepackage{txfonts}
\usepackage{hyperref}
\usepackage{lmodern}

\usepackage{tikz}
\usetikzlibrary{positioning}
\usepackage{algorithm}
\usepackage{algpseudocode}
\usepackage{statmath}
\usepackage{multirow}
\usepackage{adjustbox}

\begin{document}

\begin{frontmatter}
\title{Generalised logistic regression with vine copulas}
\cortext[cor1]{Corresponding author}
\ead{simonbb@math.uio.no / simbrant91@gmail.com}
\author{Ingrid Hob\ae k Haff}
\author{Simon Boge Brant}
\author{Haakon Bakka}
\address{}

\begin{abstract}
We propose a generalisation of the logistic regression model, that aims to account for non-linear main effects and complex
interactions, while keeping the model inherently explainable.
This is obtained by starting with log-odds that are linear in the covariates, and adding non-linear terms 
that depend on at least two covariates. More specifically, we use a generative specification of the model, consisting
of a combination of certain margins on natural exponential form, combined
with vine copulas. The estimation of the model is however based on the
discriminative likelihood, and dependencies between covariates are included 
in the model, only if they contribute significantly to the distinction between
the two classes. Further, a scheme for model selection and estimation
is presented. The methods described in this paper are implemented in the R package LogisticCopula. In order to assess the performance of our model, we ran an extensive simulation study. The results from the study, as well as from 
a couple of examples on real data, showed that our model performs 
at least as well as natural competitors, especially in the presence of
non-linearities and complex interactions, even when $n$ is not large compared to $p$.
\end{abstract}
\end{frontmatter}

\section{Introduction}
The logistic regression model is ubiquitous in discriminative modelling,
where the interest is a conditional model for a binary
outcome variable $Y$, given an observation of a $p$-dimensional covariate
vector $\mathbf{X}$. The logistic regression model in its simplest form is given by
\begin{align}
\log\left(\frac{P(Y=1\vert \mathbf{X}=\mathbf{x})}{P(Y=0\vert \mathbf{X}=\mathbf{x})}\right) = 
\beta_0 + \mathbf{x}^t\boldsymbol{\beta},\label{eq:logistic}    
\end{align}
i.e., the model for the log-odds ratio is linear. The model parameters are 
then estimated using the conditional likelihood of $Y|\mathbf{X}=\mathbf{x}$.
This model also has the advantage that it is inherently interpretable. For instance, 
the decision to accept or deny can be explained to a layperson applying for
a bank loan through how the prediction depends on whether a weighed sum of factors 
such as their age, level of education, number of children, etc., exceeds a certain 
threshold, i.e. the decision boundary. This is important particularly when used for 
applications that are affected by regulations on transparency, such as the General Data 
Protection Regulation (GDPR) of the EU, which states that any data subject (a person, a company, 
etc.) has the right to `an explanation of the decision reached after [algorithmic] assessment' 
\citep{goodman2016eu}.

The logistic regression model is a typical example of a discriminative 
framework, where the parameters are estimated by maximising the conditional likelihood of 
$Y|\mathbf{X}=\mathbf{x}$. An alternative is to use generative models for discrimination. The model
for $Y$ given $\mathbf{X}=\mathbf{x}$ is then specified \textit{indirectly} via the conditional 
distributions of $\mathbf{X}|Y=y$ for $y=0,1$ and the marginal probability $\pi_Y = P(Y = 1)$, as
\begin{equation}
P(Y = 1 \vert \mathbf{x}) = \left(\pi_Yp(\mathbf{x}\vert Y=1)\right)/
\left((1-\pi_Y)p(\mathbf{x}\vert Y=0) + \pi_Yp(\mathbf{x}\vert Y=1)\right).
\label{eq:genmod}	
\end{equation}
If one models $\mathbf{X}|Y=y$ with the multivariate Gaussian distribution, 
the result is a linear or quadratic discriminative model, whereas the Naive
Bayes model is defined by the assumption that the $X_{i}$s are conditionally 
independent given $Y$. In this setting, the parameters are estimated via the
likelihood functions of $\mathbf{X}|Y=y$ separately for the two classes.   

Restricted to certain choices of marginal distributions, Naive Bayes models lead to log-odds that 
are linear in the covariates. Hence, such a model is in some sense equivalent to a logistic 
regression model, apart from how the parameters are estimated. Several publications compare 
discriminative and  generative models that are equivalent in that sense, e.g. \cite{efron1975}, 
\cite{oneill1980}, \cite{rubinstein1997} and \cite{ng2002}. While it is true that linear 
discriminant analysis is asymptotically efficient compared to logistic regression, and that 
generative models such as a Naive Bayes model often works surprisingly well for small sample 
problems, for most applications where at least a moderate amount of training data is available, a 
discriminative model is preferable. Sometimes, it may be possible to construct a model that gives 
better predictions than the simple logistic regression. However, one should acknowledge that model 
mechanics that are comprehensible to a human have an inherent value in many applications, for 
instance for stakeholders who need to trust that the model predictions make sense - and for 
subjects who need to trust that a model-based or model-assisted decision that affects them is not 
subjecting them to unlawful discrimination. 

In this paper, we generalise the logistic regression model to account for non-linear main effects 
and interaction effects, in order to improve prediction performance, but in such a way that 
the inherent interpretability of the logistic regression model is kept. We do this by using a 
specification of the model on generative form. More specifically, we set up a model for each of 
the two classes as a combination of certain marginal distributions and a vine copula
\citep{joe1997,bedford2001probability,bedford2002vines,kurowicka2006,aas2009pair},
accounting for the dependence. The resulting model for the log odds becomes a sum of the logistic 
regression model \eqref{eq:logistic} with linear main effects and non-linear terms that involve 
two covariates or more. The idea is that the logistic regression part will provide most of the 
explanation of a particular outcome, whereas the added, more complex non-linear terms improve the 
predictions, similar to the idea behind \cite{glad2001} for linear regression. This is akin to interpretable machine learning, which focuses on creating models that are in themselves more 
understandable to a human, as opposed to explainable artificial intelligence, that attempts to 
'explain' a decision made by a model using for instance simple(r) approximations of the model  
\citep{rudin2022interpretable}. We think that in applications where explainability is valuable
and the simple linear log-odds provides an insufficiently good fit, our way of constructing 
models is more responsible, compared to employing for instance a combination of boosted tree models, or deep neural nets, and an exogenous explanation, since there must be a discrepancy between what the model computes and the explanation of the model, and therefore cases where the explanation is wrong, or else the entire model could be replaced by its explanation \citep{rudin2019stop}. 
 
Specifically, we construct our model starting from a Naive Bayes model resulting in linear log 
odds, and then add vine copula terms 
\citep{joe1997,bedford2001probability,bedford2002vines,kurowicka2006,aas2009pair} which introduce 
additional non-linear effects and interaction terms in the form of between - covariate 
dependence, conditional on the response. The resulting model for the log odds ratio becomes a sum 
of the logistic regression model (\ref{eq:logistic}) with linear main effects and some non-linear 
terms that involve two covariates or more. While we construct the form of our model based on a 
generative model specification, the parameters are found by maximising the discriminative 
likelihood of $Y\vert \mathbf{X}=\mathbf{x}.$ Other models that could be seen in this context are 
\cite{bien2013} and \cite{lim2015}, that add pairwise interactions as products of covariates 
using group-lasso. Compared to those, our model is more flexible, as it also allows more complex 
interactions, both in terms of functional form and order of interaction. Anecdotally, for one of 
the data examples, our method resulted in a model that had fewer interaction terms than the group-
lasso method, but yielded similar predictions out of sample. 

Our model can also be seen as an extension of a Naive Bayes model, by including dependence in the 
form of a vine copula, which has already been attempted, see for instance \cite{vogelaere2020}. 
A major difference in the framework we propose is it uses the 
discriminative likelihood-based estimators and not the generative ones. Further, dependence between certain 
covariates is only included when it differs sufficiently in the two classes. Hence, the purpose 
is not to model the distribution of the covariates in the two classes as well as possible,
but to make the best possible model to discriminate between the two classes.

The paper is organised as follows. Section \ref{sec:mod_ext} describes the proposed extension of 
the logistic regression model. The methods for estimation and model selection are presented in 
Section \ref{sec:mod_sel_and_est}. These first parts of the paper focus on continuous covariates, 
whereas the necessary modifications in order to include discrete covariates are given in Section
\ref{sec:discrete}. In Section \ref{sec:sim_study}, we assess the characteristics of the proposed 
model and compare it to related alternatives, and in Section \ref{sec:real_data}, we illustrate 
it on two real data examples. Finally, we make some concluding remarks in Section 
\ref{sec:concluding_remarks}.

\section{Extending the logistic regression model}\label{sec:mod_ext}
As mentioned earlier, our aim is to extend the logistic regression model in
its simplest form via a generative model specification. We will here assume
that all covariates $X_{j}$ are continuous, but our model can easily be
adapted to also handle discrete covariates, as discussed in Section
\ref{sec:discrete}.

We start by specifying a Naive Bayes type model, i.e., assuming conditional 
independence between the covariates $\mathbf{X}$ given $Y$, where the 
marginal distributions of the covariates are such that the resulting log-odds 
ratio is linear in $\mathbf{X}$ as in \eqref{eq:logistic}. Using \eqref{eq:genmod} 
with $p(\mathbf{x}\vert Y=y)=\prod_{j=1}^{p}p(x_{j}\vert Y=y)$, the log-odds ratio
becomes
\begin{align*}
\log\left(\frac{P(Y=1\vert\mathbf{x})}{P(Y=0\vert\mathbf{x})}\right) &=
\log\left(\frac{\pi_Y}{1 - \pi_Y}\right)
+ \sum_{j=1}^p\log\left(\frac{p\left(x_j\vert Y = 1\right)}
{p\left(x_j\vert Y=0\right)}\right).
\end{align*}
This means that the marginal distributions $p(x_{j}\vert Y=y)$ need to be
such that $\log(p\left(x_j\vert Y = 1\right)/p\left(x_j\vert Y=0\right)) = a_{j}+b_{j}x_{j}$
for some constants $a_{j}$ and $b_{j}$. This is fulfilled by
members of the natural exponential family \citep{morris1982natural,cox1989}, and we therefore choose to model the covariates as
$\left[X_{j}|Y=y\right] \sim N(\mu_{y,j},\sigma_{j}^{2})$, i.e $p\left(x_j\vert Y = y\right) = 1/(\sqrt{2\pi}\sigma_{j})\exp\left(-1/(2\sigma_{j}^{2})(x_{j}-\mu_{y,j})^{2}\right)$,
so that
\begin{equation*}
\log\left(\frac{p(x_j\vert Y=1)}{p(x_j\vert Y=0)}\right) = \frac{1}{2}\left(\frac{\mu_{0, j}^2 - \mu_{1, j}^2}{\sigma_j^2}\right)+\left(\frac{\mu_{1, j} - \mu_{0, j}}{\sigma_j^2}\right)x_j.
\end{equation*}
The resulting log-odds is of the form \eqref{eq:logistic} with coefficients
\begin{align}
\begin{split}	
  \beta_0 &= \log\left(\frac{\pi_{Y}}{1 - \pi_{Y}}\right) + \sum_{j = 1}^{p}\frac{1}{2}\left(\frac{\mu_{0, j}^2 - \mu_{1, j}^2}{\sigma_{j}^{2}}\right)\\
 \beta_{j}  & =(\mu_{1, j} - \mu_{0, j})/\sigma_{j}^2, \quad j = 1, \ldots, p.
\end{split}
\label{eqn:coeff}
\end{align}
Note that alternative marginal distributions for continuous covariates could also be used, such as an exponential distribution with a class-dependent parameter for covariates with support on the positive real numbers.

\subsection{Extension with vine copulas}

Our aim is to extend the above logistic regression model \eqref{eq:logistic}
to account for non-linear main effects, as well as potentially complex interaction
effects in the log odds. In order to achieve that, we use the generative representation
\eqref{eq:genmod} of the model, introducing conditional dependence between
(some of) the covariates in the form of a vine copula, but estimate the model parameters by maximising the likelihood of 
$Y|\mathbf{X}=\mathbf{x}$.

\subsubsection{Vine copulas}\label{subsec:vines}

A copula $C$ may be used to describe and analyse multivariate distributions, and 
is a multivariate distribution function whose univariate marginal distributions are all 
uniform. It allows to separate the distribution into two parts, a set of univariate marginal 
distributions, and the dependence structure. Indeed, Sklar's theorem \citep{sklar1959fonctions} 
states that if the variables $X_1, \dots, X_d$ are all continuous (which will be the case in 
the parts of the model where we introduce copulas), the joint probability density function (pdf) 
 is given by 
$$
f(x_1, \dots, x_d) = f_1(x_1)\cdot \ldots\cdot f_d(x_d)\cdot c(F_1(x_1), \dots, F_d(x_d)),
$$
where $F_1,\dots, F_d$ and $f_1, \dots, f_d$ are the marginal cumulative distribution
functions (cdf) and pdfs, respectively, and 
$c(u_1, \dots, u_d) = \partial^{d} C(u_1, \dots, u_p)/(\partial u_1\ldots \partial u_d)$ ia the density
of the unique copula $C$ of $X_1, \dots, X_d$.

A very flexible, yet computationally tractable family of copulas are vine copulas
\citep{joe1996families, bedford2001probability, bedford2002vines,kurowicka2006,
aas2009pair}. These models combine bivariate components, pair-copulas, into a 
multivariate copula, and are incredibly flexible, as all the involved pair-copulas can be
selected completely freely \citep{niko-2008}. 

Vine copulas consist of a sequence of trees $T_{i}$ with nodes $N_{i}$ and edges 
$E_{i}$, for $i=1,...,d-1$, where each edge corresponds to one of the pair-copulas that 
are the building blocks of the joint, multivariate copula. They satisfy the
following conditions \citep{vinebook}:
\begin{enumerate}
\item $T_1$ has nodes $N_1=\{1,...,d\}$ and edges $E_1$.
\item For $i=2,..., d-1$, $T_i$ has nodes $N_i=E_{i-1}$.
\item Proximity condition: two edges in tree $T_i$ can only be joined by an edge in
tree $T_{i+1}$ if they share a common node in $T_i$.
\end{enumerate}
Let $N_{ik}$ and $N_{il}$ be two nodes that are joined by the edge $e$ in $T_{i}$. Then, 
the proximity condition states that $N_{ij}$ and $N_{ik}$ must share all but one node, i.e.
$N_{ij}=\{j(e),D(e)\}$ and $N_{ik}=\{k(e),D(e)\}$, where $D(e)$ are the nodes
they have in common, called the \textit{conditioning set}, and $j(e)$ and $k(e)$ the ones
that are not shared, denoted the {\it conditioned nodes}. The edge $e$ is then associated
with the bivariate copula $C_{j(e),k(e)|D(e)}$.

The density of a vine copula for a d-dimensional random vector $\mathbf{X}$ with
marginal distributions $F_{1},\ldots,F_{d}$ is given by a product of $d(d-1)/2$ bivariate copula densities, and may
be written as
\begin{small}
\begin{align}
c_{1\dots d}(F_{1}(x_{1}),\ldots,&F_{d}(x_{d})) = 
\prod_{t=1}^{d-1}
\prod_{e\in E_t}c_{j(e),k(e)\vert D(e)}\left(u_{j(e)\vert D(e)},
u_{k(e)\vert D(e)}
\right),\label{eqn:rvine}
\end{align}
\end{small}
where $u_{j\vert D} = F(x_j\vert \mathbf{x}_D),$ and 
$\mathbf{X}_{D(e)}$ is the subvector of $\mathbf{X}$ determined by the indices
$D(e)$. An example 
of a vine copula in dimension $5$ is shown in Figure \ref{fig:fiveRvine}, which has density 
\begin{small}
\begin{align*}
  &c_{1\ldots 5} = c_{12}\cdot c_{13}\cdot c_{14}\cdot c_{35}\cdot c_{34|1}\cdot c_{23|1}\cdot c_{15|3}\cdot c_{24|13}\cdot c_{25|13}\cdot c_{45|123},
\end{align*}  
\end{small}
where arguments are omitted for simplicity. In the first tree, these arguments are the 
univariate marginal distributions, and in subsequent trees, they are conditional distributions, 
where the number of variables conditioned on is equal to the tree number minus one. In 
vine copulas, these conditional distributions
are easily computed based on pair-copulas from earlier trees, due to the proximity condition (see for instance (2) in the supplementary material). For a more 
comprehensive introduction to vine copulas, consult for instance \cite{czado2019}.

\begin{small}
\begin{figure}[t]
  \centering
\begin{tikzpicture}[
    roundnode/.style={circle, draw=green!60, fill=green!5, very thick,
      minimum size=7mm},
]
\node[roundnode]      (var1)                              {$1$};
\node[roundnode]        (var2)       [below=of var1] {$2$};
\node[roundnode]      (var3)       [right=of var1] {$3$};
\node[roundnode]        (var4)       [left=of var1] {$4$};
\node[roundnode]        (var5)       [right=of var3] {$5$};


\draw[-] (var4.east) -- (var1.west) node[anchor=north east]{$14$};
\draw[-] (var2.north) -- (var1.south) node[anchor=north west]{$12$};
\draw[-] (var1.east) -- (var3.west) node[anchor=north east]{$13$};
\draw[-] (var3.east) -- (var5.west) node[anchor=north east]{$35$};
\end{tikzpicture}

\hspace{1cm}

\begin{tikzpicture}[
    roundnode/.style={circle, draw=green!60, fill=green!5, very thick,
      minimum size=7mm},
]
\node[roundnode]      (var1)                              {$13$};
\node[roundnode]        (var2)       [left=of var1] {$14$};
\node[roundnode]      (var3)       [right=of var1] {$12$};
\node[roundnode]      (var4)       [below=of var1] {$35$};
\draw[-] (var2.east) -- (var1.west) node[anchor=north east]{$34|1$};
\draw[-] (var1.east) -- (var3.west) node[anchor=north east]{$23|1$};
\draw[-] (var4.north) -- (var1.south) node[anchor=north west]{$15|3$};
\end{tikzpicture}

\hspace{1cm}

\begin{tikzpicture}[
    roundnode/.style={circle, draw=green!60, fill=green!5, very thick,
      minimum size=7mm},
]
\node[roundnode]      (var1)                 {$23|1$};
\node[roundnode]        (var2)   [left=of var1] {$34|1$};
\node[roundnode]        (var3)   [right=of var1] {$15|3$};
\draw[-] (var2.east) -- (var1.west) node[anchor=north east]{$24|13$};
\draw[-] (var1.east) -- (var3.west) node[anchor=north east]{$25|13$};
\end{tikzpicture}

\hspace{1cm}

\begin{tikzpicture}[
    roundnode/.style={circle, draw=green!60, fill=green!5, very thick,
      minimum size=7mm},
]
(0,0)\node[roundnode]      (var1)                 {$24|13$};
(0,2)\node[roundnode]        (var2)   [right=of var1] {$25|13$};
\draw[-] (var1.east) -- (var2.west) (1.2,-0.4)node{$45|123$};
\end{tikzpicture}
\caption{Example of a vine copula for five variables.
  \label{fig:fiveRvine}}
\end{figure}
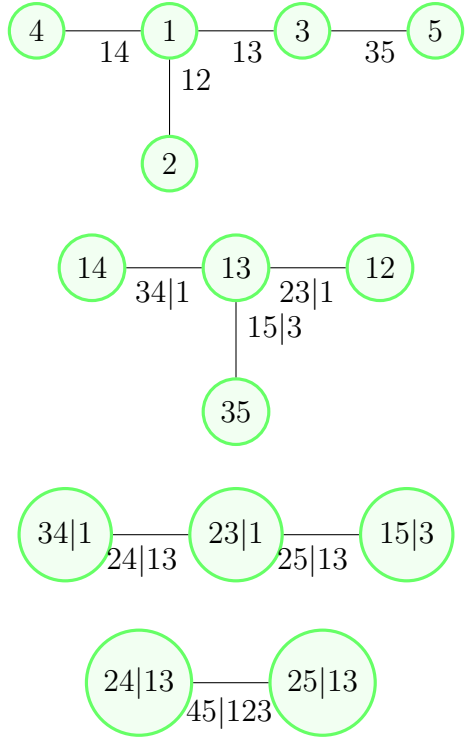

\end{small}

\subsubsection{Extension of logistic regression}\label{subsec:ext}

In order to take non-linearities and possibly complex interaction effects
into account, while maintaining an interpretable model, we allow some of
the covariates to be conditionally dependent given the class $Y$, whereas
the remaining covariates are conditionally independent. Now, the joint pdf
for $\mathbf{X}|Y=y$ becomes
\begin{align*}
    p(\mathbf{x}\vert Y = y) =  \prod_{j=1}^{p}p(x_j\vert y) \cdot c^{y}(F_{1|y}(x_1\vert y), F_{2|y}(x_2\vert y), \dots, F_{p|y}(x_p\vert y)),
\end{align*}
where $c^{y}$ is the copula density of the covariates in class $Y=y$. Letting $\boldsymbol{\mu}_{y}=(\mu_{y,1},\ldots,\mu_{y,p})$,
$\boldsymbol{\sigma}=(\sigma_{1},\ldots,\sigma_{p})$ and
$\beta_{j}$, $j=0,\ldots,p$, be as in \eqref{eqn:coeff}, we 
can express the log-odds as an extension of
\eqref{eq:logistic}:
\begin{align}
\begin{split}
\log\left(\frac{P(Y = 1 \vert \mathbf{X} = \mathbf{x})}{P(Y = 0 \vert \mathbf{X}=\mathbf{x})}\right) = &\beta_{0}+\sum_{j=1}^{p}\beta_{j}x_j+ g(x_{1},\ldots,x_{p};\boldsymbol{\mu}_{0},\boldsymbol{\mu}_{1},\boldsymbol{\sigma},\boldsymbol{\theta}_{0},\boldsymbol{\theta}_{1}),
\end{split}
\label{eqn:ext.mod}
\end{align}

where $\boldsymbol{\theta}_{0}$ and $\boldsymbol{\theta}_{1}$ are the parameters of the copulas $c^{0}$ and $c^{1}$ and $g(\cdot)$ is given by the difference in the logarithm of their densities.
The idea is then to let $c^{0}$ and $c^{1}$ be vine copulas, meaning that $c^{i}$, $i=0,1$ are given by \eqref{eqn:rvine},
resulting in
\begin{small}
\begin{align}
\begin{split}
g(x_{1},\ldots,x_{p};\boldsymbol{\mu}_{0},\boldsymbol{\mu}_{1},\boldsymbol{\sigma},\boldsymbol{\theta}_{0},\boldsymbol{\theta}_{1})=
\sum_{t=1}^{p-1}
&\sum_{e\in E_t}\left(\log\left(c_{j(e),k(e)\vert D(e)}^{1}\left(
u_{j(e)\vert D(e)}^1,
u_{k(e)\vert D(e)}^1);\boldsymbol{\theta}_{j(e),k(e)\vert D(e)}^{1}\right)\right)\right.\\
& \ \ \ \ -\left.\log\left(c_{,j(e),k(e)\vert D(e)}^{0}\left(
u_{j(e)\vert D(e)}^0,
u_{k(e)\vert D(e)}^0;\boldsymbol{\theta}_{j(e),k(e)\vert D(e)}^{0}\right)\right)\right).\label{eqn:g.func}
\end{split}	
\end{align}
\end{small}
Here, $u^y_{j(e)\vert D(e)} = F_{j(e)\vert D(e)}^{y},$ and $F_{j(e)\vert D(e)}^{y}$ is the cdf of 
$[X_{j(e)}|\mathbf{X}_{D(e)}=\mathbf{x}_{D(e)},Y=y]$, that
depends on the marginal parameters $\boldsymbol{\mu}_{y}$ and
$\boldsymbol{\sigma}$ as
$F_{j(e)}^{i}(x_{j(e)})=F_{j(e)|y}(x_{j(e)}|y=i)=\Phi((x_{j(e)}-\mu_{i,j(e)})/\sigma_{j(e)})$ when
$D(e)$ is empty, and also on copulas from earlier levels when it
is not empty. Further, $c_{j(e),k(e)\vert D(e)}^{y}$ is the copula density linking
$X_{j(e)}$ and $X_{k(e)}$, given $\mathbf{X}_{D(e)}=\mathbf{x}_{D(e)}$ and
$Y=y$, and the pair-copula parameters $\boldsymbol{\theta}_{j(e),k(e)\vert D(e)}^{y}$
constitute $\boldsymbol{\theta}_{y}$. The terms of $g$ above are then important only if 
the difference between the two log-copula densities 
$c_{j(e),k(e)\vert D(e)}^{1}$ and $c_{j(e),k(e)\vert D(e)}^{0}$ is
of a certain size. The point now is to keep only the copula pairs where this difference really contributes to the discrimination between the two classes, while setting the remaining copula densities to $1$, corresponding to (conditional)
independence between $X_{j(e)}$ and $X_{k(e)}$ given $\mathbf{X}_{D(e)}=\mathbf{x}_{D(e)}$ in
both classes.

As an example, assume that the function $g$ is given by
\begin{small}
\begin{align*}
&\log\left(c_{12}^{1}\right)-
\log\left(c_{12}^{0}\right)+\log\left(c_{13}^{1}\right)-\log\left(c_{13}^{1}\right)+\log\left(c_{14}^{1}\right)-\log\left(c_{14}^{0}\right)\\
+&\log\left(c_{34|1}^{1}\right)-\log\left(c_{34|1}^{0}\right)+\log\left(c_{23|1}^{1}\right)-\log\left(c_{23|1}^{0}\right)+\log\left(c_{24|13}^{1}\right)-\log\left(c_{24|13}^{0}\right),
\end{align*}
\end{small}
where the parameters and function arguments are omitted to simplify the notation.
This corresponds to $c^{0}$ and $c^{1}$ being given by the vine in Figure \ref{fig:fiveRvine}, 
but where the copulas $c_{35}$, $c_{15|3}$, $c_{25|13}$ and $c_{45|123}$ are set
to independence in both classes. If all $12$ copulas in the above expression are Gaussian, the resulting 
model \eqref{eqn:ext.mod} contains pairwise interactions between all the pairs constituting 
the conditioned sets of the above copulas, i.e. $(1,2)$, $(1,3)$, $(1,4)$, $(3,4)$, $(2,3)$ 
and $(2,4)$. This is a quadratic discriminative 
model, with different mean vectors in the two classes, but equal covariance 
matrices, except for the listed pairs. For other copula types, the introduction
of a copula pair generally results in an interaction of the order given by total
number of variables involved in the copulas, in addition to all the lower order
sub-interactions, e.g. the pair $(c_{34|1}^{0},c_{34|1}^{1})$
introduces a three-way interaction between $X_{1}$, $X_{3}$ and $X_{4}$, as well
as all three two-way interactions between them.

Further, the inclusion of a copula pair also changes the main effects by 
introducing non-linearities, as well as the intercept $\beta_{0}$. For 
instance if both $c_{12}^{0}$ and $c_{12}^{1}$ are Gaussian, the resulting 
contribution to the model is
\begin{align*}
\log\left(c_{12}^{1}\right)-
\log\left(c_{12}^{0}\right)=a_{0}+a_{11}x_{1}+a_{21}x_{2}&+a_{12}x_{1}^{2}+a_{22}x_{2}^{2}+a_{3}x_{1}x_{2},
\end{align*}
with
\begin{small}
\begin{align*}
a_{0} = & -\frac{1}{2}\log\left(\frac{1-(\theta_{12}^{1})^{2}}{1-(\theta_{12}^{0})^{2}}\right)+\frac{1}{2}\sum_{y=0}^{1}(-1)^{y}\frac{(\theta_{12}^{y})^{2}}{1-(\theta_{12}^{y})^{2}}\left(\frac{\mu_{y,1}^{2}}{\sigma_{1}^{2}}+\frac{\mu_{y,2}^{2}}{\sigma_{2}^{2}}-2\frac{\mu_{y,1}\mu_{y,2}}{\sigma_{1}\sigma_{2}}\right)\\
a_{i1} = & \frac{1}{\sigma_{i}^{2}}\sum_{y=0}^{1}(-1)^{y}\frac{(\theta_{12}^{y})^{2}}{1-(\theta_{12}^{y})^{2}}\left(\mu_{y,i}-\frac{\mu_{y,3-i}}{\theta_{12}^{y}}\frac{\sigma_{i}}{\sigma_{3-i}}\right), \ i=1,2\\
a_{i2} = & -\frac{1}{2\sigma_{i}^{2}}\left(\frac{(\theta_{12}^{1})^{2}}{1-(\theta_{12}^{1})^{2}}-\frac{(\theta_{12}^{0})^{2}}{1-(\theta_{12}^{0})^{2}}\right), \ i=1,2, \ a_{3} = \frac{1}{\sigma_{1}\sigma_{2}}\left(\frac{\theta_{12}^{1}}{1-(\theta_{12}^{1})^{2}}-\frac{\theta_{12}^{0}}{1-(\theta_{12}^{0})^{2}}\right).  
\end{align*}
\end{small}
We see that the quadratic and the interaction effects disappear if
$\theta_{12}^{0}=\theta_{12}^{1}$, meaning that the copula parameter is the
same in both classes. This is a particular trait of the bivariate normal
distribution, which is the resulting distribution when combining a Gaussian copula with Gaussian margins. However, non-linearities and interactions will in general appear
when introducing copula terms in the model, even if the copulas are the same in
both classes, as long as the margins, i.e. the means $\mu_{1,i}$
and $\mu_{0,i}$, are different.
  
The sizes of the copula log-differences depend both on whether the two copulas
are different and on whether the copula arguments are different in the
two classes. A difference in the copulas could be due to a different copula family, different parameters, or both. Differences in the copula arguments between the
two classes may be caused by differences in the corresponding marginal 
distributions in the two classes, by different copulas on lower levels, or both.
A difference in the margins means that the corresponding main effect is present
in the model, whereas a difference in lower level copulas means that lower order
interaction effects between the involved variables are present in the model.

\section{Estimation and model selection}\label{sec:mod_sel_and_est}
Below, we describe how the model parameters are estimated in Section \ref{subsec:est}
and propose a method for selecting the model in Section \ref{subsec:mod_sel}.

\subsection{Estimation of model parameters}\label{subsec:est}
Each time a couple of copula terms are added to the model, all model parameters are 
re-estimated by maximum likelihood as described below. This is essential in order to
avoid unnecessarily including copula terms that do not contribute to the
discrimination.

The full parameter vector is 
$\boldsymbol{\eta}=(\boldsymbol{\beta}^{T},\boldsymbol{\theta}^{T})^{T}$, where
$\boldsymbol{\beta}=(\beta_{0},\ldots,\beta_{p})^{T}$
is the vector of coefficients from the logistic regression without
interactions, and $\boldsymbol{\theta}=(\boldsymbol{\theta}_{0}^{T},\boldsymbol{\theta}_{1}^{T})^{T}$
is the vector of parameters of the log-copula densities. Let the observed data be
$y_{1},\ldots,y_{n}$ and $\textbf{x}_{1},\ldots,\textbf{x}_{n}$, $\textbf{x}_{i}$
being the $p$-dimensional covariate vector corresponding to the response $y_{i}$.
The likelihood function is given by
\begin{small}
\begin{align}
\begin{split}l(\boldsymbol{\beta},\boldsymbol{\theta};y_{1},\ldots,y_{n},\textbf{x}_{1},\ldots,\textbf{x}_{n}) = &\sum_{i=1}^{n}y_{i}\log\left(\frac{p_{i}(\boldsymbol{\beta},\boldsymbol{\theta},\textbf{x}_{i})}{1-p_{i}(\boldsymbol{\beta},\boldsymbol{\theta},\textbf{x}_{i})}\right)+\sum_{i=1}^{n}\log\left(1-p_{i}(\boldsymbol{\beta},\boldsymbol{\theta},\textbf{x}_{i})\right)\\
= &\sum_{i=1}^{n}y_{i}(1,\textbf{x}_{i})^t\boldsymbol{\beta}+\sum_{i=1}^{n}y_{i}g(\textbf{x}_{i};\boldsymbol{\mu}_{0}(\boldsymbol{\beta}),\boldsymbol{\mu}_{1}(\boldsymbol{\beta}),\boldsymbol{\sigma}(\boldsymbol{\beta}),\boldsymbol{\theta})\\
-&\sum_{i=1}^{n}\log\left(1+\exp\left\{(1,\textbf{x}_{i})^t\boldsymbol{\beta}+g(\textbf{x}_{i};\boldsymbol{\mu}_{0}(\boldsymbol{\beta}),\boldsymbol{\mu}_{1}(\boldsymbol{\beta}),\boldsymbol{\sigma}(\boldsymbol{\beta}),\boldsymbol{\theta})\right\})\right).
\label{eqn:loglik}
\end{split}
\end{align}
\end{small}
where we have plugged in the log odds ratio from \eqref{eqn:ext.mod} in the last equality.

Maximising \eqref{eqn:loglik} over the full parameter vector simultaneously might be
time-consuming and numerically challenging. However, preliminary parameter estimates
$\hat{\boldsymbol{\beta}}^{0}$ and $\hat{\boldsymbol{\theta}}^{0}$
from the model selection routine provide useful start values for a numerical
optimiser. Further, the optimisation is done with a quasi-Newton method using the 
gradients of the likelihood function with respect to the parameters. Descriptions of how to 
obtain the these gradients are given in the supplementary material.

Under the usual regularity assumptions, and in particular that the assumed model is the 
true data generating mechanism, the maximum likelihood estimator fulfils
\[
\sqrt{n}\left(\left(\hat{\boldsymbol{\beta}}^{T},\hat{\boldsymbol{\theta}}^{T}\right)^{T}-\left(\boldsymbol{\beta}^{T},\boldsymbol{\theta}^{T}\right)^{T}\right) \indist  N_{m}(\boldsymbol{0},\boldsymbol{J}^{-1}),
\]
where $m$ is the total number of parameters and 
$\boldsymbol{J}=-\mbox{E}\left(\frac{\partial^{2} l(\boldsymbol{\beta},\boldsymbol{\theta};\textbf{Y},\textbf{x})}{\partial (\boldsymbol{\beta}^{T},\boldsymbol{\theta}^{T})^{T}\partial (\boldsymbol{\beta}^{T},\boldsymbol{\theta}^{T})}\right)$
is the Fisher information matrix. This may be used to compute confidence intervals for the parameters
$\boldsymbol{\beta}$ and $\boldsymbol{\theta}$. One may also compute
confidence intervals for any smooth function $\psi(\boldsymbol{\beta},\boldsymbol{\theta})$ of
the parameters, using the delta method
\[
\sqrt{n}(\hat{\psi}-\psi(\boldsymbol{\beta},\boldsymbol{\theta}))=\sqrt{n}(\psi(\hat{\boldsymbol{\beta}},\hat{\boldsymbol{\theta}})-\psi(\boldsymbol{\beta},\boldsymbol{\theta})) \indist N(0,\sigma_{\psi}^{2})
\]
with $\sigma_{\psi}^{2} = \frac{\partial\psi(\boldsymbol{\beta},\boldsymbol{\theta})}{\partial\left(\boldsymbol{\beta}^{T},\boldsymbol{\theta}^{T}\right)}\boldsymbol{J}^{-1}\frac{\partial\psi(\boldsymbol{\beta},\boldsymbol{\theta})}{\partial\left(\boldsymbol{\beta}^{T},\boldsymbol{\theta}^{T}\right)^{T}}$.
An example of such a function is the log-odds
\[
\psi(\boldsymbol{\beta},\boldsymbol{\theta})=\log\left(\frac{p_{i}(\boldsymbol{\beta},\boldsymbol{\theta},\textbf{x}_{i})}{1-p_{i}(\boldsymbol{\beta},\boldsymbol{\theta},\textbf{x}_{i})}\right)=(1,\textbf{x}_{i})\boldsymbol{\beta}+g(\textbf{x}_{i};\boldsymbol{\mu}_{0}(\boldsymbol{\beta}),\boldsymbol{\mu}_{1}(\boldsymbol{\beta}),\boldsymbol{\sigma}(\boldsymbol{\beta}),\boldsymbol{\theta}),
\] 
with
\begin{align*}
\frac{\partial\psi(\boldsymbol{\beta},\boldsymbol{\theta})}{\partial\left(\boldsymbol{\beta}^{T},\boldsymbol{\theta}^{T}\right)^{T}} = &(1,\textbf{x}_{i})+\frac{1}{c^{1}}\frac{\partial c^{1}}{\partial\left(\boldsymbol{\beta}^{T},\boldsymbol{\theta}^{T}\right)^{T}}-\frac{1}{c^{0}}\frac{\partial c^{0}}{\partial\left(\boldsymbol{\beta}^{T},\boldsymbol{\theta}^{T}\right)^{T}},
\end{align*}
where $\frac{\partial c^{y}}{\partial\left(\boldsymbol{\beta}^{T},\boldsymbol{\theta}^{T}\right)^{T}}$
is given in expressions (2) and (3) in the supplementary material.

\subsection{Model selection}\label{subsec:mod_sel}
The first step in building the proposed model is to estimate the main effects
$\boldsymbol{\beta}$ by iteratively re-weighted least squares,
resulting in $\hat{\boldsymbol{\beta}}$. To be able to extend this model as
described in Section \ref{sec:mod_ext}, the parameter estimates must be converted to 
estimates of the parameters $\mu_{0,j}$, $\mu_{1,j}$ and $\sigma_{j}$, $j = 1, \ldots, p$, 
of the corresponding marginal distributions in the Naive Bayes model specification. 
As the number of marginal parameters is larger than the length of 
$\boldsymbol{\beta}$, there are infinitely many sets of marginal parameters that 
result in the same main effects. To find a unique solution, we choose to impose the following restrictions: $\mbox{E}(X_{j})$ should match $\bar{x}_{j}=1/n\sum_{i=1}^{n}x_{ij},$ and $\mbox{Var}(X_{j})$ should match 
$s_{j}^{2}=1/(n-1)\sum_{i=1}^{n}(x_{ij}-\bar{x}_{j})^{2}$, for $j=1,\ldots,p$. Using
\begin{small}
\begin{align*}
\mbox{E}(X_{j}) = (1-\pi_{Y})\mu_{0, j}+\pi_{Y}\mu_{1, j} \quad \mbox{Var}(X_{j}) = \sigma_j^2 + \left(\mu_{1, j} - \mu_{0, j}\right)^2
\pi_Y(1 - \pi_Y), \quad j = 1\ldots,p
\end{align*}
\end{small}
together with expression \eqref{eqn:coeff} for the coefficients, we obtain
the following equation, that must be solved for $\hat{\pi}_{Y}$
\begin{align}
\begin{split}
\log\left(\frac{\hat{\pi}_Y}{1 - \hat{\pi}_Y}\right) = &\hat{\beta}_{0}
- \sum_{j=1}^{p}\frac{1}{2}
\left(\frac{\hat{\mu}_{0, j}(\hat{\pi}_{Y})^2 - \hat{\mu}_{1, j}(\hat{\pi}_{Y})^2}{\widehat{\sigma_j^2}(\hat{\pi}_{Y})}\right) 
\end{split}
\label{eqn:py}
\end{align}
plugging in
\begin{align}
\begin{split}
\widehat{\sigma_{j}^{2}}(\hat{\pi}_{Y}) = \frac{\sqrt{1 + 4\hat{\beta}_{j}^{2}\hat{\pi}_{Y}(1 - \hat{\pi}_{Y})s_{j}^2} - 1}{2\hat{\beta}_{j}^2\hat{\pi}_{Y}(1 - \hat{\pi}_{Y})}, \
\begin{matrix}
\hat{\mu}_{0, j}(\hat{\pi}_{Y}) = &\bar{x}_{j} - \hat{\beta}_{j}\hat{\pi}_{Y}\widehat{\sigma_{j}^2}(\hat{\pi}_{Y})\\
\hat{\mu}_{1, j}(\hat{\pi}_{Y}) = &\bar{x}_{j} + \hat{\beta}_{j}(1 - \hat{\pi}_{Y})\widehat{\sigma_{j}^2}(\hat{\pi}_{Y})
\end{matrix}
\end{split}
\label{eqn:sigma.mu}
\end{align}
for $j = 1,\ldots,p$. All the other parameter estimates are subsequently obtained by 
plugging in the resulting value of $\hat{\pi}_{Y}$ into the above expressions. 

The model is then ready for expansion with vine copula terms. The structures
of the two vines $c^{0}$ and $c^{1}$ are restricted to be the same. The purpose
of this is to ensure that all terms of $\log(c^{0})$ are matched by a corresponding 
term in $\log(c^{1})$, so that the corresponding interaction effects are easier to 
interpret. However, the copula family of a given pair of variables is allowed to be
different in the two classes, for instance a Gumbel copula in class $0$ and a Clayton
in class $1$.

The procedure for extending the model starts by including the copulas of the first
trees of $c^{0}$ and $c^{1}$.  In the first step, all pairs of variables, $(x_{j},x_{k})$,
$j,k \in \{1,\ldots,p\}$ with $j \neq k$, are considered, and the pair that provides
the largest increase in the likelihood function (given by \eqref{eqn:loglik}) when
the copulas $c_{jk}^{0}$ and $c_{jk}^{1}$ are included, is chosen. To reduce the
computational burden in this model selection procedure, all copulas are assumed to be
Gaussian at this stage. Once a pair has been selected, other one-parameter copula
families are also considered, to find the ones for the two classes that increase the
likelihood the most. Then, the parameters are re-estimated as described in Section
\ref{subsec:est}.

After the first step, only pairs that involve a variable already included
in the vine copula are considered, under the additional restriction that no cycles
should be formed in the tree. For instance, if the pairs $(1,2)$ and $(2,3)$ have been
included, $(1,3)$ cannot. 

The process of adding copulas to the first tree is continued while the likelihood ratio
is larger than a pre-specified threshold $e^{\tau}$, which is equivalent to 
\begin{align*}
l^{new}(\hat{\boldsymbol{\beta}}^{new},\hat{\boldsymbol{\theta}}^{new};y_{1},\ldots,y_{n},\textbf{x}_{1},\ldots,\textbf{x}_{n})-l^{cur}(\hat{\boldsymbol{\beta}}^{cur},\hat{\boldsymbol{\theta}}^{cur};y_{1},\ldots,y_{n},\textbf{x}_{1},\ldots,\textbf{x}_{n}) \geq \tau,
\end{align*}
where the log-likelihood function is given by \eqref{eqn:loglik} in Section \ref{subsec:est}.
The value of $\tau$ should be chosen so as to avoid over-fitting, without being too
strict. The procedure may stop adding pairs before the tree has the maximum number
$d-1$ of edges. In this case, the remaining edges are assumed to be independence copulas.

After this, copulas are added to the second tree, again until the log-likelihood difference 
is no longer larger than $\tau$, after which copulas are added to the third tree and so on.
Again, all copulas are assumed to be Gaussian while searching for the
next pair of copulas to be included in the model. Once it is selected, the optimal copula
family is found for each of the two classes, and the model parameters are re-estimated.
From the second tree, the pair-copulas must fulfil the proximity condition for $c^{0}$ and
$c^{1}$ to be legal vine copulas (see Section \ref{subsec:vines}). This 
drastically reduces the number of pairs that must be considered for inclusion in the
model. In addition, it is required that the two copulas from the previous tree, needed
to compute the conditional distributions that are copula arguments must be present
in the model for a given pair of copulas to be part of the model. Hence, $c_{13|2}$ may
only be included if $c_{12}$ and $c_{23}$ are. This corresponds to an
assumption of strong hierarchy\citep{bien2013} for higher order interactions that is mainly made for
computational purposes. Extensions to weak or no hierachies are straightforward.

The maximum number $K$ of trees in the final vines is also the maximum order of the 
interactions in the model, and is a parameter that should be chosen. However, this
only means that $c^{0}$ and $c^{1}$ will have at most $K$ trees, but the selection
procedure may stop before that number is reached, and copulas in the remaining
trees of the vine are then set to independence, which is called truncation
\citep{brechmann2012}. The full model selection procedure is described in Algorithm
1 in the supplementary material.

\section{Including discrete covariates}\label{sec:discrete}
In order to include discrete covariates in the model, we need
marginal distributions for the generative model specification.
As explained in Section \ref{sec:mod_ext}, their forms must be
such that the resulting log-odds ratio is linear in the
covariates, which is fulfilled by the members of the exponential
family. Hence, categorical covariates with $d$ categories are
represented as $d-1$ binary dummy variables, and each dummy variable
$X_{j}$ is then modelled as $\left[X_{j}|Y=y\right] \sim \text{Bernoulli}(\rho_{y,j})$, i.e. $p(x_j\vert Y=y) = (1 - \rho_{y,j})^{1 - x_j}\rho_{y,j}^{x_j}$. Further, counting covariates are modelled as $\left[X_{j}|Y=y\right] \sim \text{Poisson}(\lambda_{y,j})$, i.e. $p(x_j\vert Y=y) = \lambda_{y,j}^{x_{j}}/x_{j}! \cdot\exp(-\lambda_{y,j})$.

Now, let $\mathcal{C}$, $\mathcal{D}$ and $\mathcal{E}$ denote the (possibly empty) sets
of indices corresponding to continuous, binary and counting covariates, respectively.
Then, the log-odds ratio is still of the
form \eqref{eq:logistic} with coefficients
\begin{align}
\begin{split}	
\beta_0 &= \log\left(\frac{\pi_Y}{1 - \pi_Y}\right)
+ \sum_{j\in \mathcal{C}}\frac{1}{2}
\left(\frac{\mu_{0, j}^2 - \mu_{1, j}^2}{\sigma_j^2}\right) +
\sum_{j\in\mathcal{D}}\log(\rho_{1, j}/\rho_{0, j}) +
\sum_{j\in\mathcal{E}}\left(\lambda_{0,j}-\lambda_{1,j}\right),\\
\beta_j  &=
\begin{cases}
(\mu_{1, j} - \mu_{0, j})/\sigma_j^2, &\text{ if } j \in \mathcal{C}\\
\log\left(\frac{\rho_{1,j}(1 - \rho_{0, j})}{(1 - \rho_{1, j})\rho_{0,j}}\right), &\text{ if } j \in \mathcal{D}\\
\log(\lambda_{1, j}/\lambda_{0, j}), &\text{ if } j \in \mathcal{E}
\end{cases}.
\end{split}
\label{eqn:coeff.discr}
\end{align}

For the model selection procedure, the first estimates $\hat{\boldsymbol{\beta}}$ of the main effects
must now be converted to estimates of the parameters $\mu_{0,j}$, $\mu_{1,j}$ and $\sigma_{j}$,
$j \in \mathcal{C}$, $\rho_{0,j}$ and $\rho_{1,j}$, $j \in \mathcal{D}$, and $\lambda_{0,j}$ and 
$\lambda_{1,j}$, $j \in \mathcal{E}$, of the corresponding marginal distributions, matching $\mbox{E}(X_{j})$ to
$\bar{x}_{j}$ for $j \in \mathcal{C}, \mathcal{D}, \mathcal{E}$ 
and $\mbox{Var}(X_{j})$ to $s_{j}^{2}$, which results in the following equation that must be solved
for  $\hat{\pi}_{Y}$
\begin{align}
\begin{split}
\log\left(\frac{\hat{\pi}_Y}{1 - \hat{\pi}_Y}\right) = &\hat{\beta}_{0}
- \sum_{j\in \mathcal{C}}\frac{1}{2}
\left(\frac{\hat{\mu}_{0, j}(\hat{\pi}_{Y})^2 - \hat{\mu}_{1, j}(\hat{\pi}_{Y})^2}{\widehat{\sigma_j^2}(\hat{\pi}_{Y})}\right) -
\sum_{j\in\mathcal{D}}\log\left(\frac{\hat{\rho}_{1,j}(\hat{\pi}_{Y})(1 - \hat{\rho}_{0, j}(\hat{\pi}_{Y}))}{(1 - \hat{\rho}_{1, j}(\hat{\pi}_{Y}))\hat{\rho}_{0,j}(\hat{\pi}_{Y})}\right)\\ 
&-
\sum_{j\in\mathcal{E}}\left(\hat{\lambda}_{0,j}(\hat{\pi}_{Y})-\hat{\lambda}_{1,j}(\hat{\pi}_{Y})\right)
\end{split}
\label{eqn:py.discr}
\end{align}
plugging in \eqref{eqn:sigma.mu} for $\widehat{\sigma_{j}^{2}}(\hat{\pi}_{Y})$, $\hat{\mu}_{0, j}(\hat{\pi}_{Y})$ and $\hat{\mu}_{1, j}(\hat{\pi}_{Y})$, for $j \in \mathcal{C}$,
\begin{align}
\begin{split}
\hat{\rho}_{0,j}(\hat{\pi}_{Y}) = &\frac{1 + \left(1 - e^{\hat{\beta}_{j}}\right)\left(\bar{x}_{j}-\hat{\pi}_{Y}\right)-\sqrt{\left(1 + \left(1 - e^{\hat{\beta}_{j}}\right)\left(\bar{x}_{j}-\hat{\pi}_{Y}\right)\right)^{2}-4\left(1 - e^{\hat{\beta}_{j}}\right)\left(1 - \hat{\pi}_Y\right)\bar{x}_{j}}}{2\left(1 - e^{\hat{\beta}_{j}}\right)\left(1 - \hat{\pi}_Y\right)}\\
\hat{\rho}_{1, j}(\hat{\pi}_{Y}) = &\frac{e^{\hat{\beta}_{j}}\hat{\rho}_{0, j}}{1 - \left(1 - e^{\hat{\beta}_{j}}\right)\hat{\rho}_{0, j}}	
\end{split}
\label{eqn:rho.discr}
\end{align}
for $j \in \mathcal{D}$ and
\begin{align}
\begin{split}
\hat{\lambda}_{0, j}(\hat{\pi}_{Y}) = \frac{\bar{x}_{j}}{1-(1-e^{\hat{\beta}_{j}})\hat{\pi}_{Y}}, \ \hat{\lambda}_{1, j}(\hat{\pi}_{Y}) = \hat{\lambda}_{0, j}(\hat{\pi}_{Y}) e^{\hat{\beta}_{j}} 
\end{split}
\label{eqn:lambda.discr}
\end{align}
for $j \in \mathcal{E}$. The rest of the model selection procedure remains the same,
as the discrete covariates are not considered for interaction effects.

In principle, one could include also the discrete covariates in the vine copulas
of expression \eqref{eqn:g.func}, with densities replaced by ratios of probability
mass functions. However, even if model selection and estimation methods for
vine copulas with discrete components have been proposed (see for instance
\cite{panagiotelis2012} and \cite{panagiotelis2017}), it is much more
challenging to do inference with discrete data, and in particular binary
variables \citep{neslehova2007}. This is a topic for further work, and beyond the
scope of this paper.

\section{Simulation study}\label{sec:sim_study}
In order to assess how our method performs, we have conducted a simulation study. The first part
(Section \ref{subsec:perf1}) concerns the performance under different sizes of the
data set and the choice of threshold for the model selection (see Section \ref{subsec:mod_sel}),
whereas the second part (Section \ref{subsec:perf2}) is a comparison to other methods.

\subsection{Simulation setup}\label{subsec:setup}

We have simulated data from four different generative models, all with 
$Y\sim Bernoulli(0.5)$ to ensure balanced data. The first models are of the form 
$\mathbf{X}|Y=y \sim N(\mathbf{\mu}_{y},\mathbf{\Sigma}_{y})$, for $y=0,1$, with
$\mathbf{\Sigma}_{y}=\mathbf{D}_{y}\mathbf{R}_{y}\mathbf{D}_{y}$, $D_{y}$ being a diagonal 
matrix with the vector $\boldsymbol{\sigma}_{y}$ of standard deviations on the diagonal. The 
distinction between the two models is that in Model 1, the standard deviations are different in the 
two classes, whereas in Model 2, they are the same, i.e.,
$\boldsymbol{\sigma}_{0}=\boldsymbol{\sigma}_{1}$. Further, the correlation matrices 
$\mathbf{R}_{0}$ and $\mathbf{R}_{1}$ are different in some elements, which results in a 
discriminative model on logistic regression form, with some two-way interactions in addition to 
quadratic main effects. This means that Model 2 is a special case of our model, but since we
assume equal standard deviations, Model 1 is not.

In the two other models we have simulated from, i.e. Models 3 and 4, $\mathbf{X}|Y=y$ has the 
same marginal distributions as in Models 1 and 2, respectively, but the dependence is modelled by 
a vine copula composed of Gumbel and Clayton copulas, whereas the dependence structures of 
Models 1 and 2 can be seen as a vine copula composed of Gaussian copulas. The vine copulas of
Models 3 and 4 are such that the structure and copula types are the same in the two classes, but the 
parameters of some of the copulas in the first three trees are different. This results in some 
two-way, three-way and four-way interactions, as well as non-linear main-effects, that are not just 
quadratic. 

For the number $p$ of covariates in $\mathbf{X}$, we have used the values $8$, $14$ and $20$,
and for the size of the training set $n=500$, $1000$ and $2000$ independent observations. The
parameters in the two classes have been set such that the interactions and non-linearities are
quite important, so that a simple logistic regression model is insufficient. Moreover, we
have for each setting simulated $100$ data sets of size $n$, as well as $100$ independent 
test sets of size $10000$. Further, we consider the copula types Gaussian, Gumbel and Clayton
for our model, in order to allow both symmetric and asymmetric dependencies.

\subsection{Effect of the amount of data and the model selection threshold}\label{subsec:perf1}

As explained in Section \ref{subsec:mod_sel}, the model selection procedure requires a
threshold parameter $\tau$ for deciding which copula components should be included in the
model. To study the effect of this threshold on the performance of our model, we have 
fitted each simulated training set using each of the $6$ values $1$, $1.5$, $2$, 
$2+(\log(n)-2)/3$, $2+2(\log(n)-2)/3$ and $\log(n)$, where $2$ and $\log(n)$ correspond
to using the AIC and BIC, respectively. Subsequently, we have computed the out-of-sample
area under the curve (AUC) and log-likelihood on the test sets. 

\begin{figure}[p]
\includegraphics[width=0.45\textwidth]{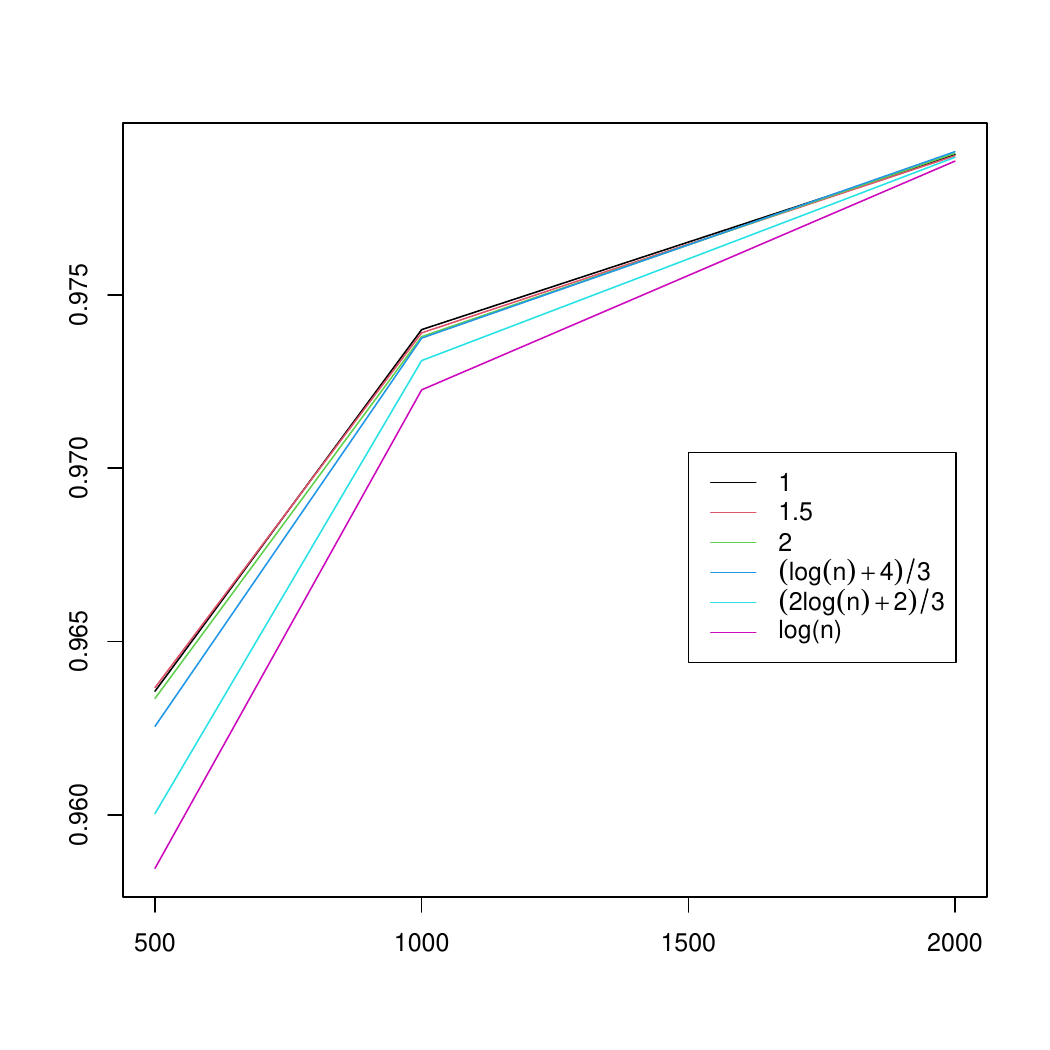}
\includegraphics[width=0.45\textwidth]{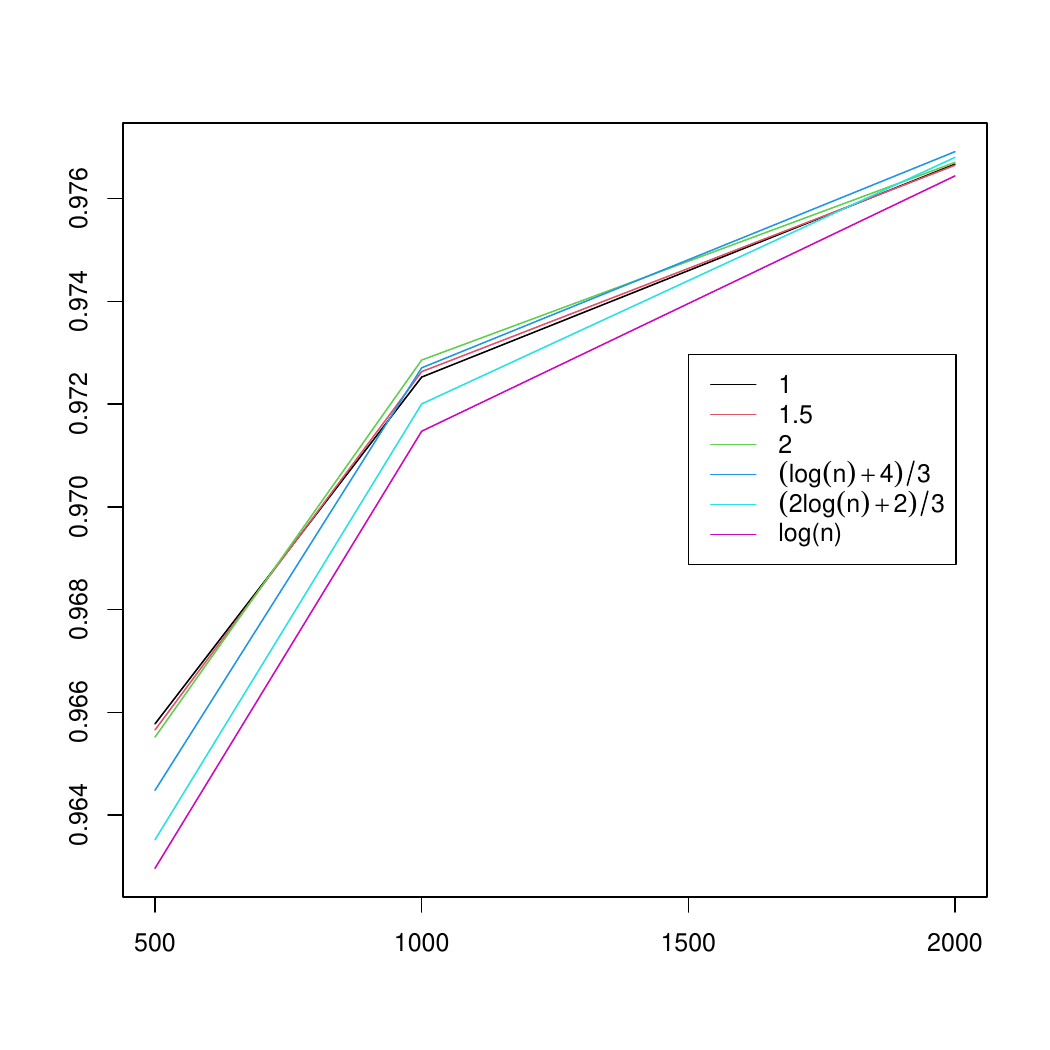}\\
\includegraphics[width=0.45\textwidth]{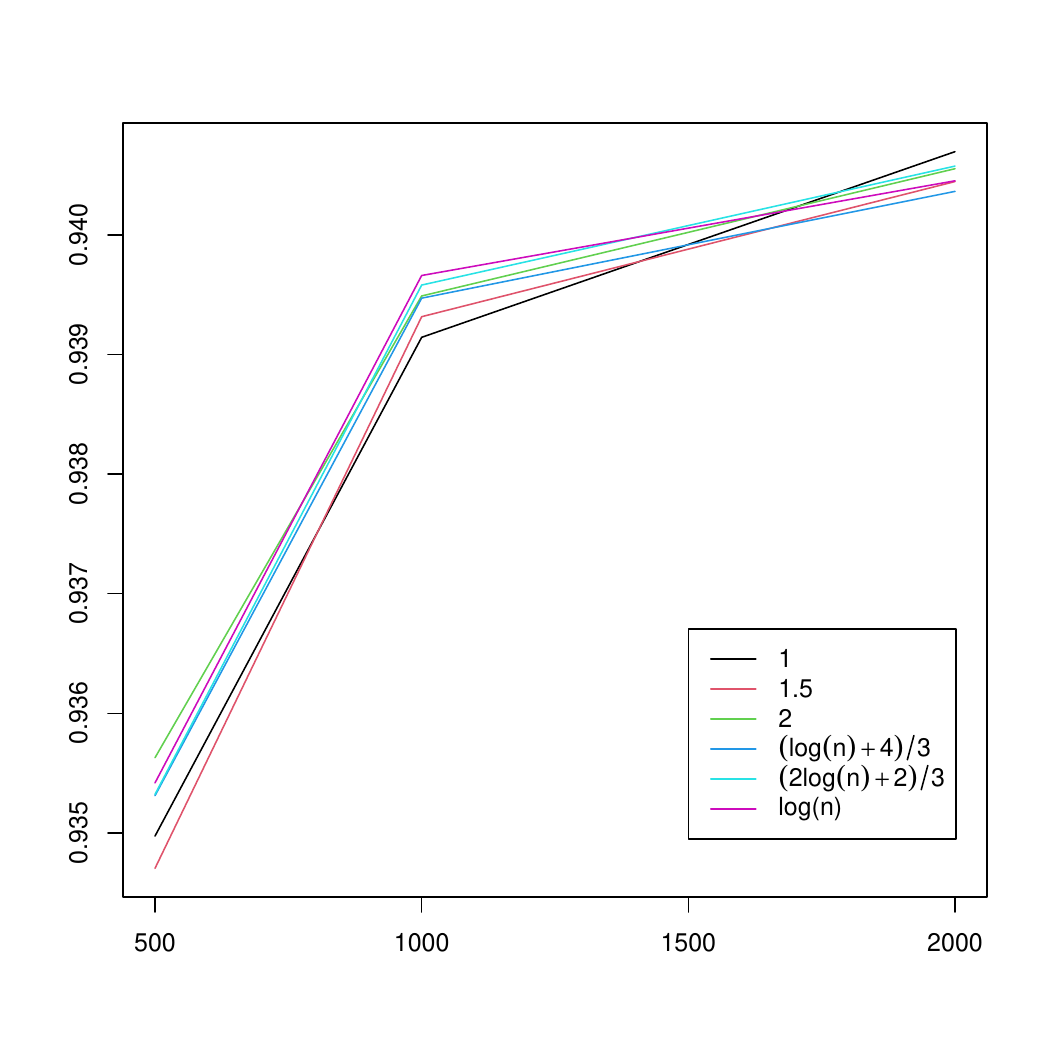}
\includegraphics[width=0.45\textwidth]{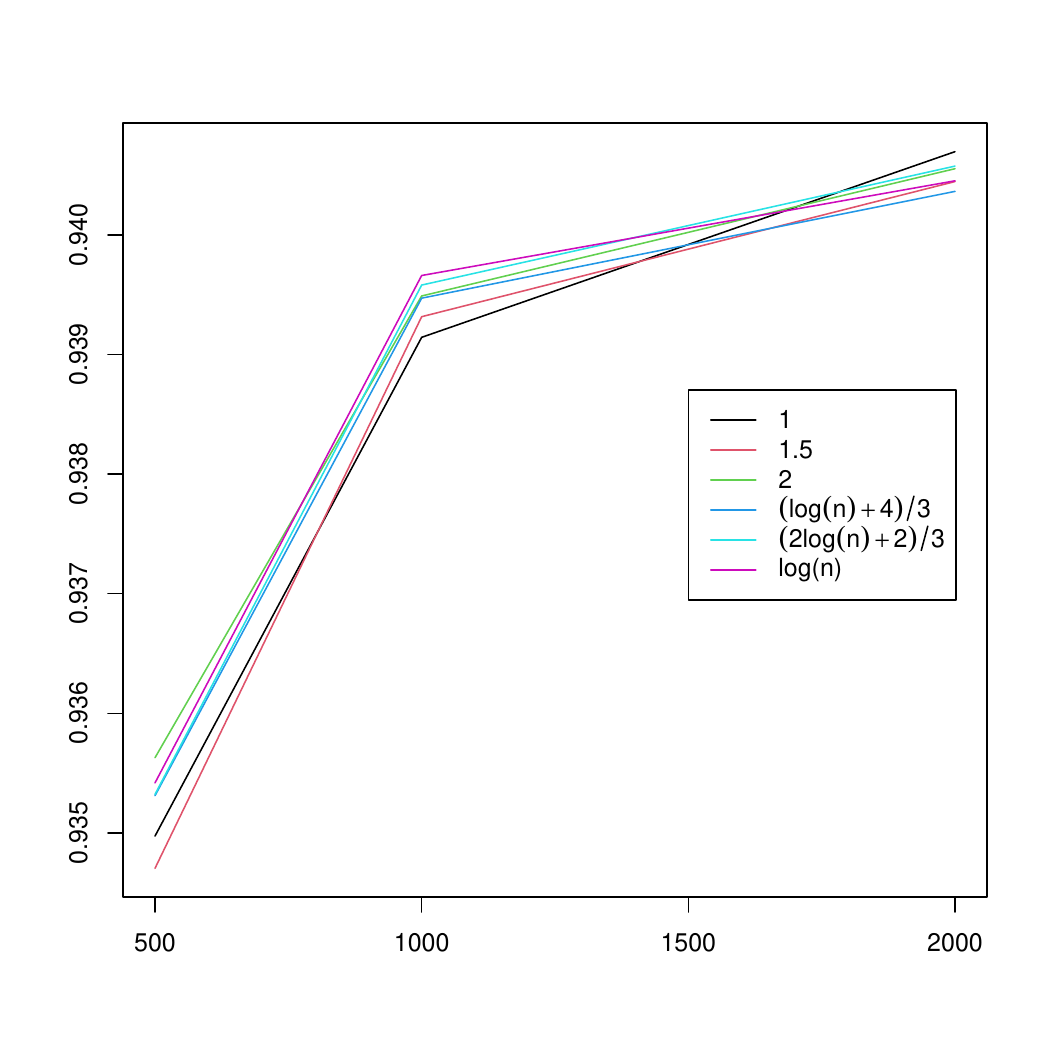}
\caption{Out-of-sample AUC from fitting our model to simulations from Models 1 to 4 with $p=8$ covariates as a function of the size $n$ of the training set for different values of the threshold $\tau$. The values have been averaged over $100$ data sets. \label{fig:auc_8}}
\end{figure}

The corresponding out-of-sample AUCs for the four models with $p=8$ are shown in Figure 
\ref{fig:auc_8}. First, note that the y-axis covers a rather small range, in order to bring out 
the differences, but that the AUCs generally are high for all sample sizes and threshold values, 
indicating that our model has very good prediction performance. In comparison, the optimal AUCs 
for the test sets, obtained using the true models, are $0.97-0.98$. As expected, the AUCs
increase with the sample size, but are already high for the smallest training sample size
$n=500$, indicating that our method does not necessarily require a large amount of data to perform
well. Further, the different thresholds do not affect the performance in terms of the AUC that
much, especially for larger $n$. 

\begin{figure}[p]
\includegraphics[width=0.45\textwidth]{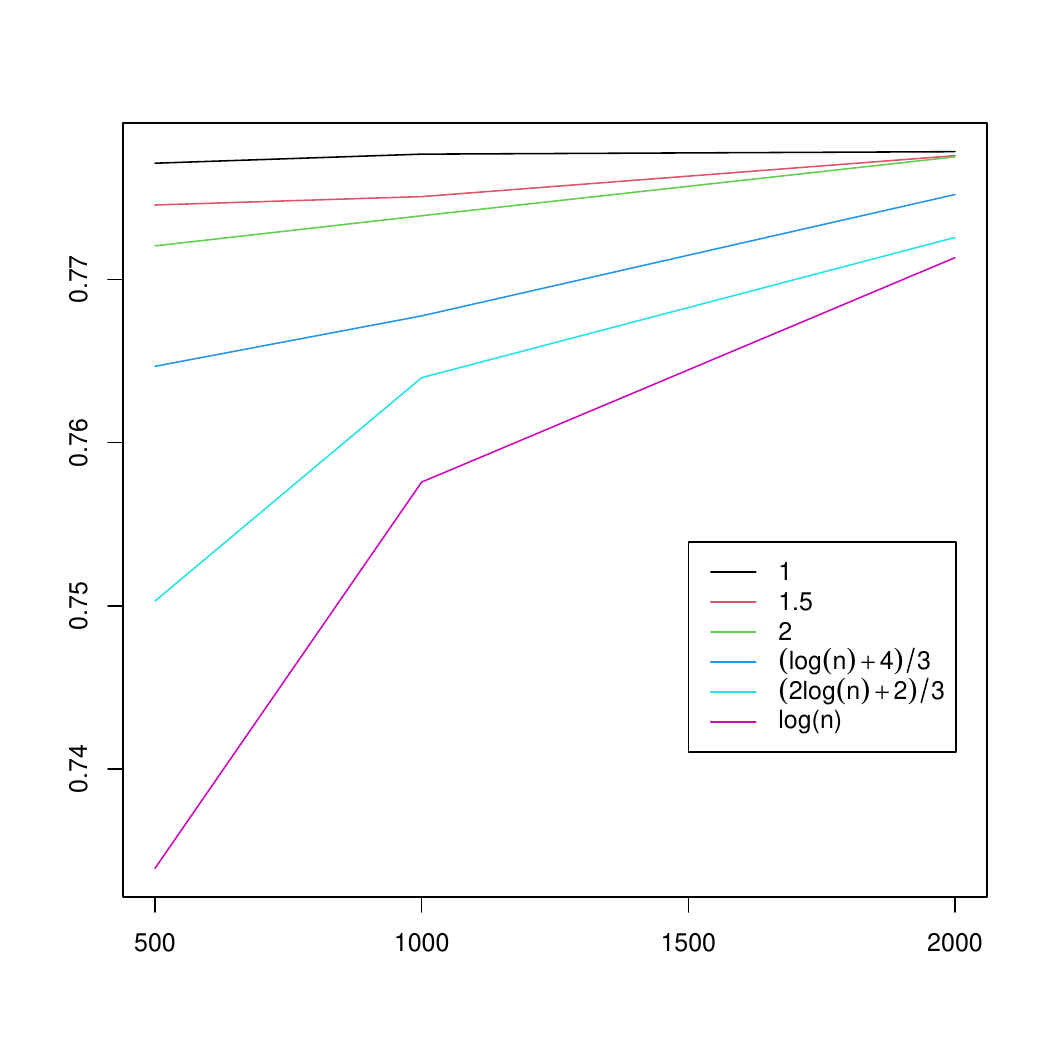}
\includegraphics[width=0.45\textwidth]{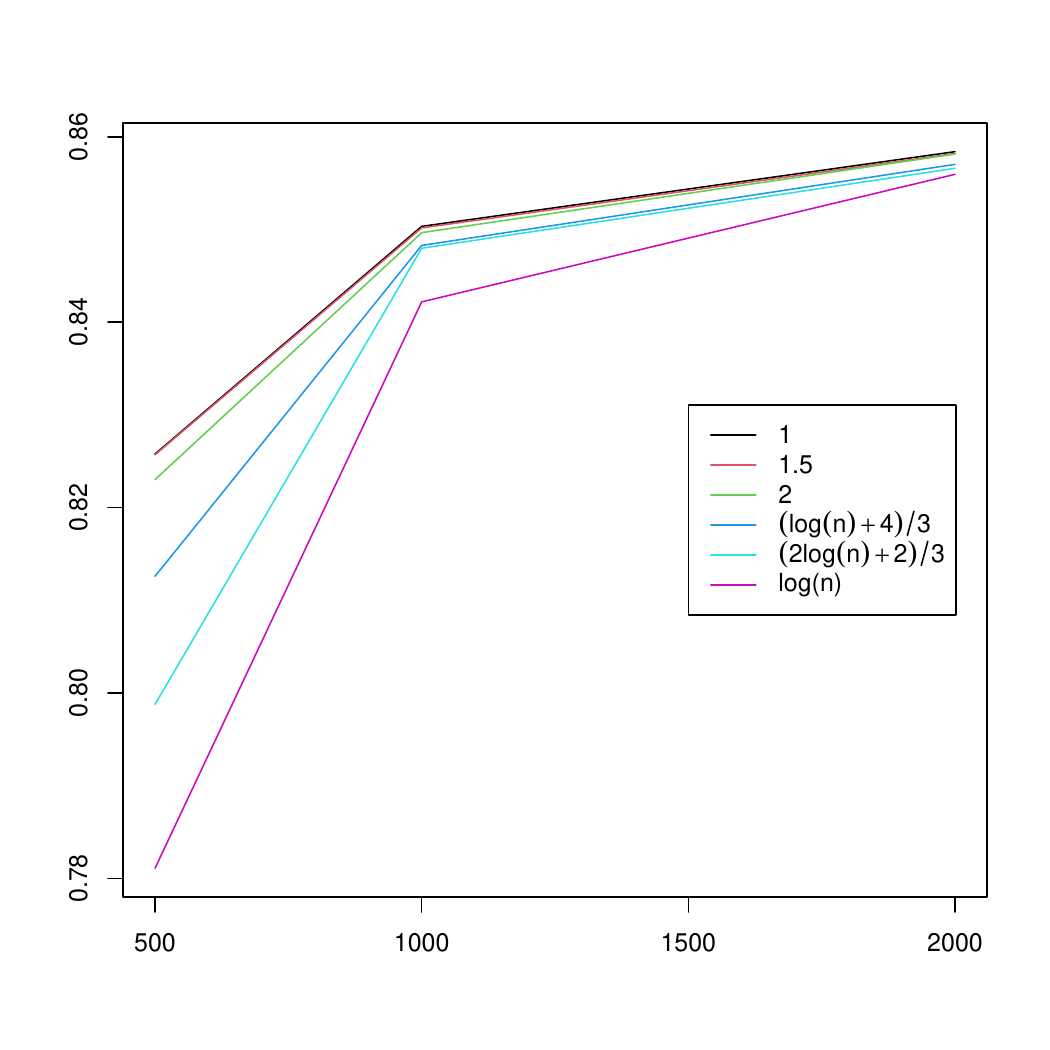}\\
\includegraphics[width=0.45\textwidth]{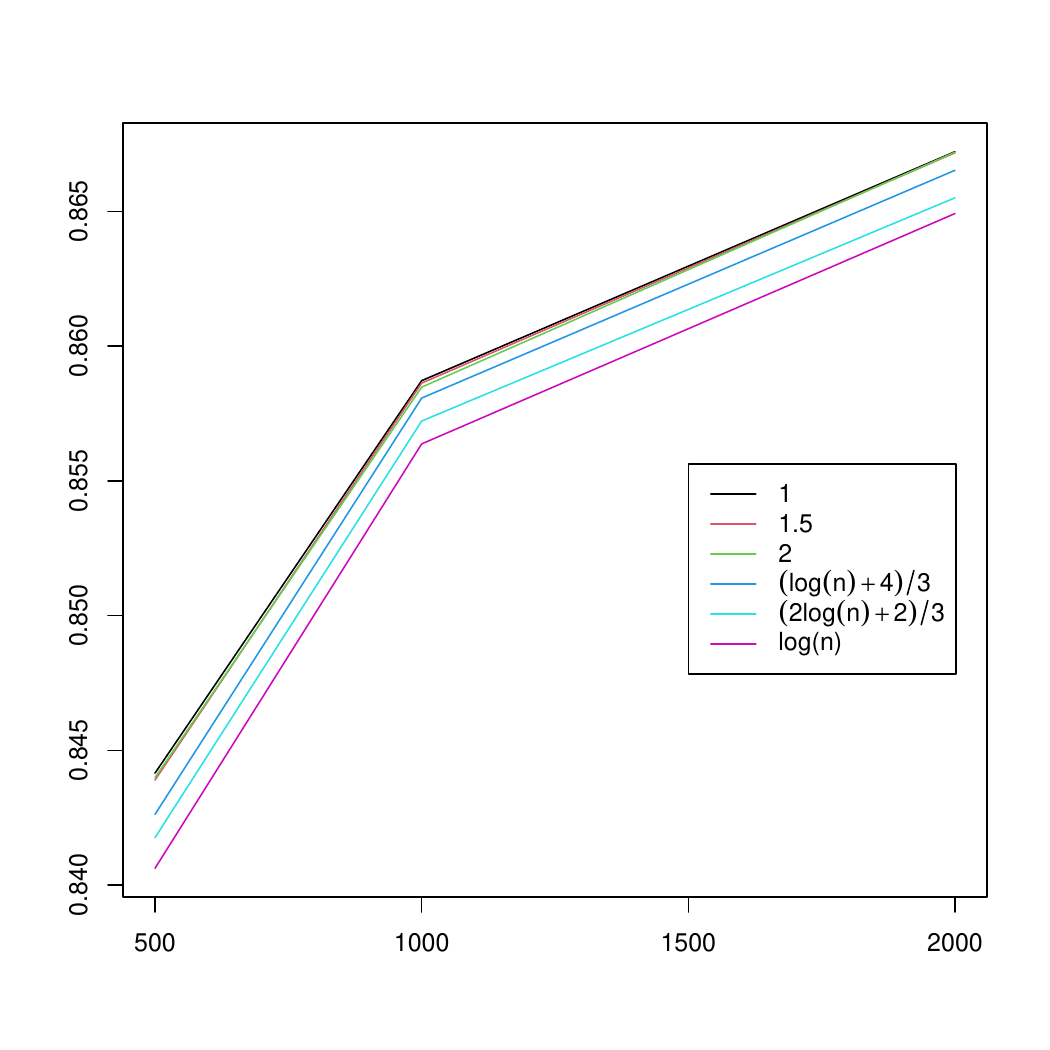}
\includegraphics[width=0.45\textwidth]{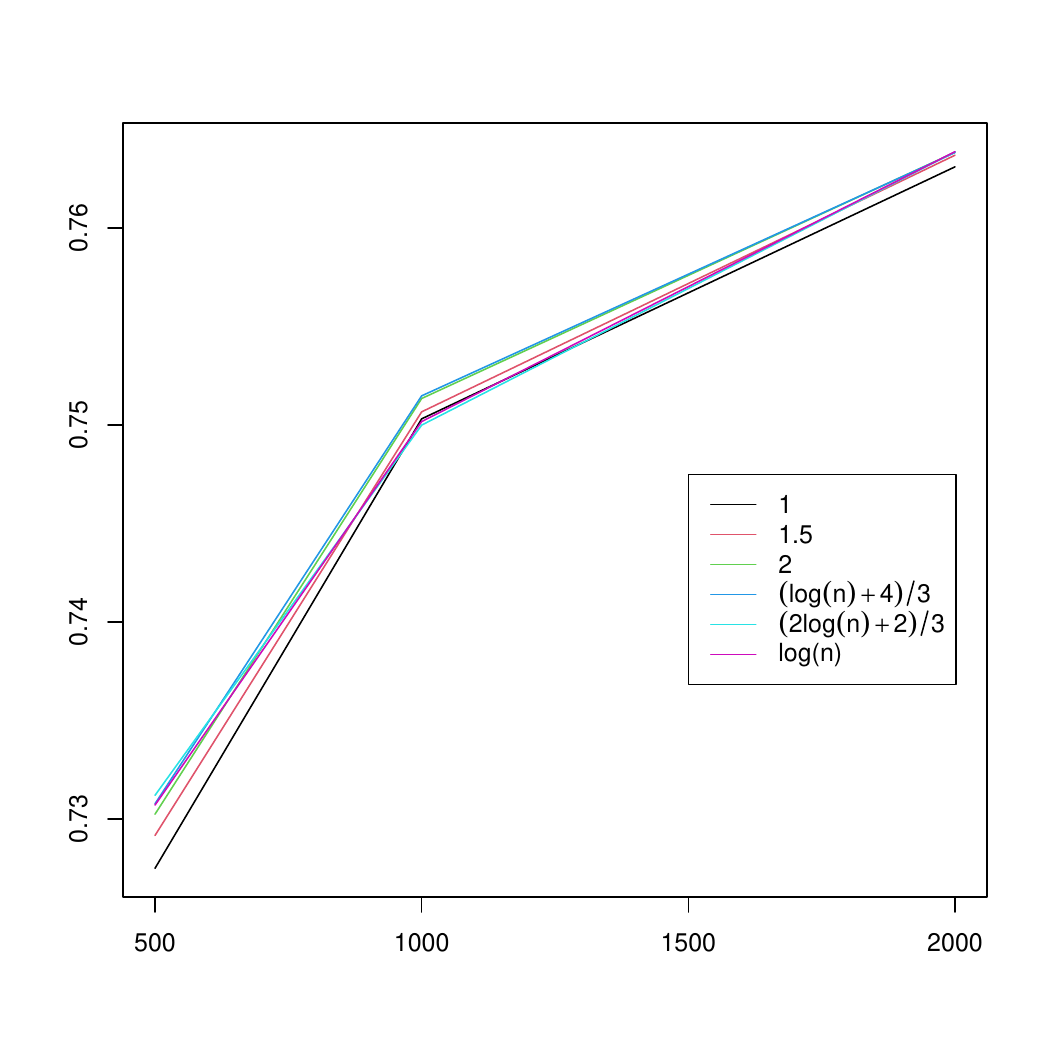}
\caption{Out-of-sample AUC from fitting our model to simulations from Models 1 to 4 with $p=8$ covariates and weaker non-linearities and interaction effects as a function of the size $n$ of the training set for different values of the threshold $\tau$. The values have been averaged over $100$ data sets. \label{fig:auc_2_8}}
\end{figure}

We repeated the above simulations but from models with $p=14$ and $p=20$ covariates.
The results in terms of the out-of-sample AUC are quite similar to the ones for $p=8$
with very high AUCs even for the lowest training set size and small differences for 
different values of the threshold, and are shown in the supplementary material. We also
ran simulations from the four models with $p=8$ and weaker non-linearities and interaction
effects, resulting in an optimal AUC on the test sets between $0.82$ and $0.90$. The
corresponding AUC values are given in Figure \ref{fig:auc_2_8}. The performance of our
model is again comparable, except that the choice of threshold has a larger impact for
the smallest sample size for Models 1 and 2, where values of at most $2$ seem to be
the best. Finally, corresponding plots of the log-likelihood, shown in the supplementary
material, show similar patterns.

\begin{figure}[p]
\includegraphics[width=0.45\textwidth]{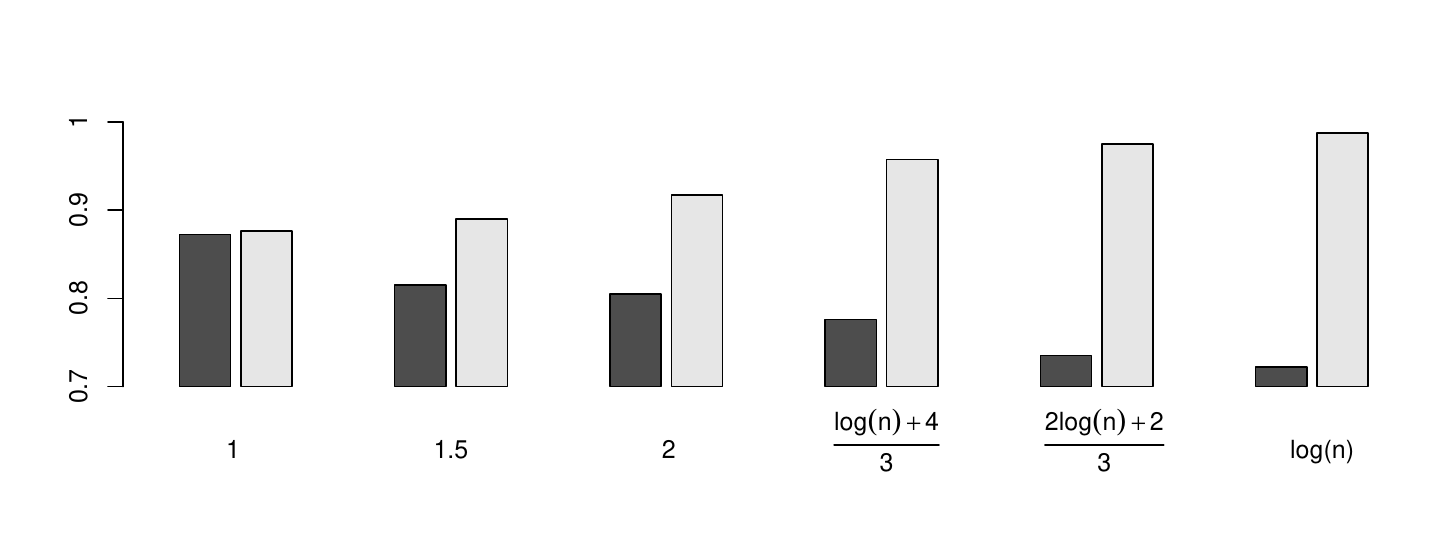}
\includegraphics[width=0.45\textwidth]{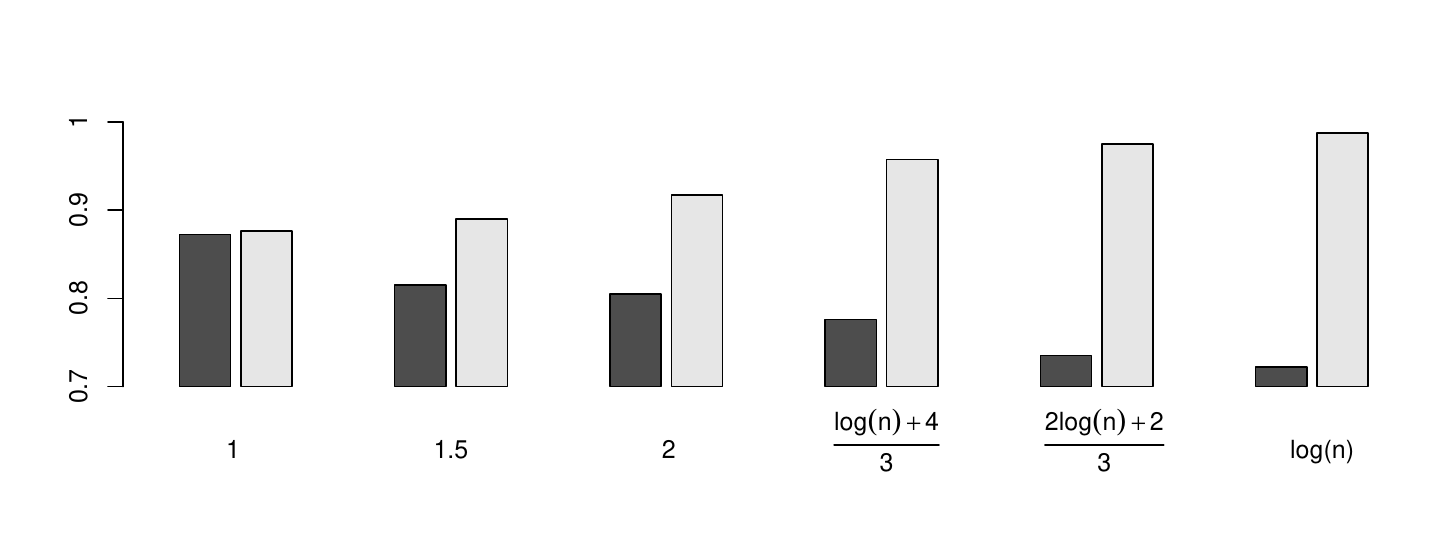}\\
\includegraphics[width=0.45\textwidth]{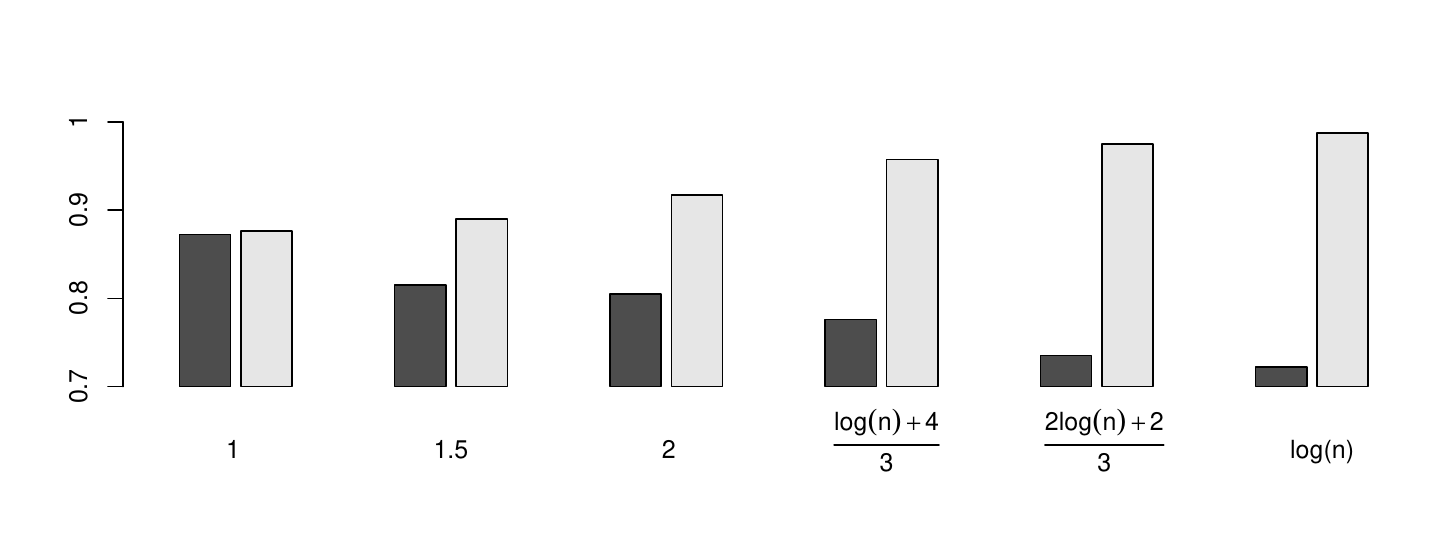}
\includegraphics[width=0.45\textwidth]{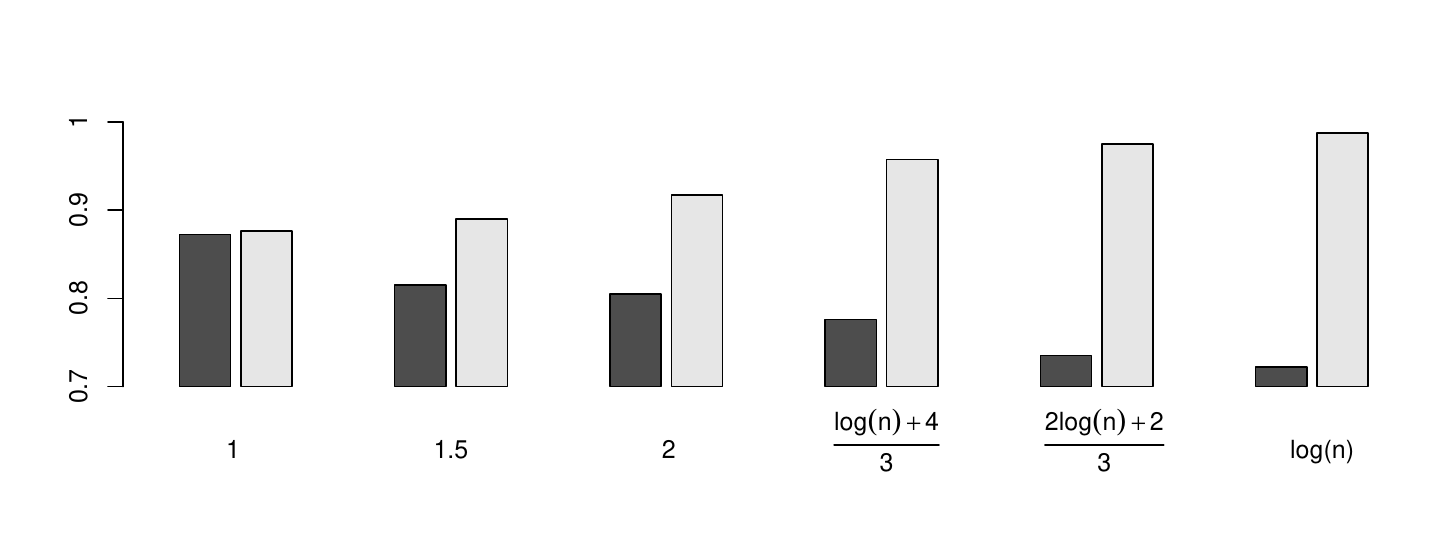}
\caption{Fraction of true interactions that are found by our model (in dark gray) and of the selected interactions that are true interactions (in light gray) when fitting our model to simulations from Models 1 to 4 with $p=8$ and $n=500$ as a function of the  $\tau$. The values have been averaged over $100$ data sets. \label{fig:found_true_int_8}}
\end{figure}

The choice of threshold may also depend on whether the interaction effects picked out
by the method are important in themselves for the purpose of interpretation. In Figure
\ref{fig:found_true_int_8} we have plotted the fraction of the interaction
effects from the data generating mechanism that are found by our method, as well as the 
fraction among the interactions selected with our method that are true interactions 
for the case with $p=8$, $n=500$ and strong non-linearities and interactions. Overall, 
we see that a large percentage of the true interactions are found by our method, and 
most of the found interactions are true interactions from the simulation
models. As expected, lower values of the threshold results in finding a larger
portion of the true interaction effects, since more copula terms are added to the model.
On the other hand, there are also more interactions included in the model, that should not
be there, but the fraction of false discoveries is low for all values of the threshold, 
indicating that the interactions picked out by the model can be trusted and interpreted.
Overall, the values $\tau=2$ and $\tau=2+(\log(n)-2)/3$ seem to give a good balance between
these two concerns. The pattern is the same for $n=1000$ and $n=2000$ (plots shown in the supplementary
material).

Generally, the threshold value does not have a huge effect on the results, but overall,
the value $\tau=2$ seems to be one of the best. On the other hand, the largest value 
$\tau=\log(n)$ also gives good results, and has the advantage of resulting in simpler models and
faster fitting. We have however used the value $\tau=2$ in the next part of the simulation study
(Section \ref{subsec:perf2}).

\subsection{Comparison to other methods}\label{subsec:perf2}

As our model (denoted ``Cop LR'' in the tables below) is an extension of the simple
logistic regression model with only linear main effects (denoted ``Lin LR''), this is a natural
reference, and thus one of the models we compare ours to. More specifically, we use the R package
\texttt{glm} to fit such a model to all the simulated data sets. As mentioned earlier, our method
may also be seen as an extension of a Naive Bayes model with specific marginal distributions.
Therefore, we also include a Naive Bayes type model, namely one with normal margins (denoted
``NB ''). We also tried such a model with kernel density estimators for the margins, but it gave
either comparable or worse results than the one with normal margins, and the corresponding 
results are therefore not shown here. In the Naive Bayes model each margin is estimated separately 
with maximum likelihood, dividing the data according to class. This model can represent quadratic main 
effects, but not interaction effects. Finally, we include a logistic regression model with linear main 
effects and pairwise interactions (denoted ``Int LR''), which is another way of extending the logistic
regression model. This model is fitted using the method by \cite{lim2015}, which is based on grouped 
lasso, and is implemented in the R package \texttt{glinternet}. Here, we use 10-fold cross validation to 
find the optimal penalty, which is the default in \texttt{glinternet}.

The results in terms of the AUC for $p=8$ for the models with strong interactions and
non-linearities are given in Table \ref{tab:auc_8}. The first thing we note is that the simple
linear model performs very poorly. Further, the two Naive Bayes models give somewhat better
results, but not nearly as good as the logistic regression model with pairwise interactions. This
means that both non-linearities and interaction effects really are essential for discriminating
between the two classes in this case. Further, we see that our method provides even better results
than the logistic regression model with pairwise interactions, especially for data from Models
3 and 4, which contain more complex interaction effects, but also for Models 1 and 2, which
only contain pairwise interactions. Results for the models with $p=14$ and $p=20$
covariates are quite similar, and are displayed in the supplementary material. So are
the log-likelihoods which also follow the same patterns.

The results for the models with weaker interactions and non-linearities are given in Table
\ref{tab:auc_8_2}. The simple linear model performs better in this case, which illustrates
that the linear part of the main effects is relatively more important. Still, our method
mostly gives better results than all the other methods and never worse, even for the
smallest training set sizes. Hence, the non-linearities and interactions do not necessarily
have to be that strong for our method to be useful.

Finally, since our model has potential to become quite complex, there may be a risk of
over-fitting, especially if the true data generating mechanism is such that the
corresponding discriminative model has log odds that are close to linear. In order to investigate this, we
also simulated data from a model on the same form as Model 2, but with all correlations
of $\mathbf{R}_{0}$ and $\mathbf{R}_{1}$ set to $0$, that we denote Model 5. The corresponding log odds of the corresponding
discriminative model only has linear main effects and no interactions. The results from fitting
our model and the competitors to data from this model with $p=8$ are given in Table
\ref{tab:auc_8_3}. We see that even in this case the results from our method are comparable to
the others. That being said, our model is much slower to fit than the others, and is mainly
useful when there really are non-linearities and complex interactions in the data.

\begin{table}[p]
\begin{tabular}{lccccc}
\hline
\hline
\textbf{Sim. mod.} & \textbf{n} & \textbf{Cop LR} & \textbf{Lin LR} & \textbf{NB} & \textbf{Int LR}\\
\hline
\multirow{3}{6em}{\textbf{Model 1}} & $500$ & 0.963 & 0.501 & 0.627 & 0.933\\
& $1000$ & 0.974 & 0.501 & 0.639 & 0.939\\
& $2000$ & 0.979 & 0.501 & 0.644 & 0.943\\
\hline
\multirow{3}{6em}{\textbf{Model 2}} & $500$ & 0.966 & 0.501 & 0.501 & 0.935\\
& $1000$ & 0.973 & 0.501 & 0.501 & 0.940\\   
& $2000$ & 0.977 & 0.501 & 0.500 & 0.943\\
\hline
\multirow{3}{6em}{\textbf{Model 3}} & $500$ & 0.936 & 0.501 & 0.622  & 0.688\\
& $1000$ & 0.939 & 0.501 & 0.635 & 0.716\\
& $2000$ & 0.941 & 0.501 & 0.641 & 0.730\\
\hline
\multirow{3}{6em}{\textbf{Model 4}} & $500$ & 0.916 & 0.501 & 0.500 & 0.701\\
& $1000$ & 0.923 & 0.501 & 0.500 & 0.721\\
& $2000$ & 0.934 & 0.502 & 0.498 & 0.732\\
\hline
\hline
\end{tabular}
\caption{AUC from fitting our model, as well as simple logisitic regression, Naive Bayes and logistic regression with pairwise interactions to simulations from Models 1 to 4 with $p=8$ covariates for different sizes $n$ of the training set. The values have been averaged over $100$ data sets. \label{tab:auc_8}}
\end{table}

\begin{table}[p]
\begin{tabular}{lccccc}
\hline
\hline
\textbf{Sim. mod.} & \textbf{n} & \textbf{Cop LR} & \textbf{Lin LR} & \textbf{NB} & \textbf{Int LR}\\
\hline
\multirow{3}{6em}{\textbf{Model 1}} & $500$ & 0.772 & 0.589 & 0.769 & 0.735\\
& $1000$ & 0.774 & 0.598 & 0.774 & 0.751\\
& $2000$ & 0.778 & 0.602 & 0.776 & 0.759\\
\hline
\multirow{3}{6em}{\textbf{Model 2}} & $500$ & 0.823 & 0.574 & 0.534  & 0.695\\
& $1000$ & 0.850 & 0.582 & 0.544 & 0.712\\ 
& $2000$ & 0.858 & 0.587 & 0.551 & 0.720\\
\hline
\multirow{3}{6em}{\textbf{Model 3}} & $500$ & 0.844 & 0.585 & 0.749 & 0.714\\
& $1000$ & 0.858 & 0.592 & 0.753 & 0.737\\
& $2000$ & 0.867 & 0.596 & 0.756 & 0.749\\
\hline
\multirow{3}{6em}{\textbf{Model 4}} & $500$ & 0.730 & 0.623 & 0.558 & 0.667\\ 
& $1000$ & 0.751 & 0.630 & 0.567 & 0.688\\ 
& $2000$ & 0.764 & 0.633 & 0.572 & 0.699\\
\hline
\hline
\end{tabular}
\caption{AUC from fitting our model, as well as simple logisitic regression, Naive Bayes and logistic regression with pairwise interactions to simulations from Models 1 to 4 with $p=8$ covariates and weaker non-linearities and interactions for different sizes $n$ of the training set. The values have been averaged over $100$ data sets. \label{tab:auc_8_2}}
\end{table}

\begin{table}[t]
\begin{tabular}{ccccc}
\hline
\hline
\textbf{n} & \textbf{Cop LR} & \textbf{Lin LR} & \textbf{NB} & \textbf{Int LR}\\
\hline
$500$ & 0.939 & 0.942 & 0.942 & 0.940\\
$1000$ & 0.942 & 0.943 & 0.943 & 0.942\\
$2000$ & 0.943 & 0.944 & 0.944 & 0.943\\
\hline
\hline
\end{tabular}
\caption{AUC from fitting our model, as well as simple logisitic regression, Naive Bayes and logistic regression with pairwise interactions to simulations from Model 5 with $p=8$ covariates for different sizes $n$ of the training set. The values have been averaged over $100$ data sets. \label{tab:auc_8_3}}
\end{table}

\section{Real data examples}\label{sec:real_data}
Next, we illustrate our method on a couple of real data sets, namely the ionosphere
(Section \ref{subsec:ionosphere}) and the smoking data (Section \ref{subsec:smoking}).

\subsection{Ionoshere data}\label{subsec:ionosphere}
First, we consider a dataset of radar returns that was discussed in a paper by \citet{sigillito1989classification}, where the objective is to classify observations as 
either `bad' or `good', so that bad returns may be removed automatically. The dataset 
contains $351$ observations of $34$ covariates, as well as the binary outcome variable, 
of which $126,$ roughly $36\%,$ have the value `bad', and the rest are `good'. Two of 
the covariates are binary, but one of them only takes the value $0$, and the other is 
always $1$ when the outcome variable is `good'. Therefore, we choose to remove them 
from the dataset, leaving $32$ continuous covariates. We split the data into a training set of 
$263$ and a test set of $88$ observations, which corresponds to a $75/25\%$ split. 
Further, we let the outcome $y$ be $1$ if the observation is `bad', and $0$ if it is `good', and 
the balance between $0$s and $1$s is approximately the same in both the training and the test 
sets as in the full data set.

We fit our model, as well as the three competitor models from the simulation study. The
corresponding results in term of the out-of-sample AUC and log-likelihood on the test 
set are given in the top two rows of Table \ref{tab:real_data}. Our model provides by far the 
best results both in terms of the AUC and the log-likelihood. The improvement over the 
simple logistic regression with linear log-odds is huge, indicating that there are important 
non-linearities and/or interaction effects for explaining the response. The fact that both the 
Naive Bayes and the logistic regression with pairwise interactions also perform much better 
than the simple model, indicate that both non-linear main effects and interactions contribute.

\begin{table}[t]
\begin{tabular}{lcccccc}
\hline
\hline
\textbf{Data set} & \textbf{Measure} & \textbf{Cop LR} & \textbf{Lin LR} & \textbf{NB} & \textbf{Int LR}\\
\hline
\multirow{2}{9em}{\textbf{Ionoshere data}} & AUC & 0.957 & 0.766 & 0.926 & 0.890\\
& log-likelihood & -21.1 & -93.7 & -105.2 & -95.7\\
\hline
\multirow{2}{9em}{\textbf{Smoking data}} & AUC & 0.839 & 0.831 & 0.792 & 0.837\\
& log-likelihood & -6476.0 & -6548.1 & -19856.7 & -6484.6\\
\hline
\hline
\end{tabular}
\caption{Out-of-sample AUC and log-likelihood from fitting our model (Cop LR), as well as simple logistic regression (Lin LR), Naive Bayes (NB) and logistic regression with pairwise interactions (Int LR), to the ionoshphere (top two rows) and the smoking data (bottom two rows).\label{tab:real_data}}
\end{table}

\subsection{Smoking data}\label{subsec:smoking}
Next, we look at the body signal of smoking data set from Kaggle \citep{smoking-prediction}, that consists of
$55962$ observations of the response, which is an indicator of whether the person
is smoking or not, and $25$ covariates, describing a set of bio-indicators, such
as age, height, weight, blood pressure and cholesterol. Out of these, $18$ are continuous. 
The covariate \textit{oral} only takes one value, and is  therefore removed. The remaining 
$6$ are categorical, more specifically binary, except \textit{Urine.protein}, which takes 
$6$ categories. A preliminary analysis indicated that the categorical covariates 
\textit{eyesight.right} and \textit{Urine.protein} have little or no effect of the smoking status, 
and were therefore also removed, the former, probably because it almost always has the same 
value as \textit{eyesight.left}, whereas very few observations were in categories $2$-$6$ for 
the latter. The remaining data consists of $18$ continuous covariates and $4$ binary ones, in 
addition to the response. Finally, we made a $75/25\%$ split of the data into a training set of
$41769$ and a test set of $13923$.

The out-of sample AUC and log-likelihood from fitting the same models as to the ionosphere 
data are shown in the last two rows of Table \ref{tab:real_data}. Our model gives the highest
AUC and log-likelihood value, though very closely followed by the logistic regression model with
pairwise interactions. The Naive Bayes model has the poorest relative performance, which may be due to the large sample size, as it tends to be comparatively better in relation to discriminative methods for smaller sample sizes \citep{ng2002}. Further, there is some improvement from using the simple logistic regression model, but not 
nearly as large as for the ionosphere data. This means that the linear part of the log odds provides the 
largest part of the explanation of the response, but better predictions may be obtained by 
including interactions and non-linearities. 

It should be noted that our fitted model includes seven copulas in the first tree of the vine copula and none in the following trees, resulting in $1+22+2\cdot 7=37$ parameters in total. 
This means that it contains seven pairwise interactions and none with higher order, more 
specifically between \textit{age} and \textit{systolic} (blood pressure), between \textit{eyesight.right} and \textit{relaxation} (blood pressure), between  \textit{systolic} and 
\textit{HDL} (cholesterol type), between \textit{HDL} and \textit{LDL} (cholesterol type), 
between \textit{LDL} and \textit{hemoglobin}, between \textit{age} and \textit{ALT} (glutamic 
oxaloacetic transaminase type) and between \textit{weight.kg} and \textit{serum.creatinine}.
On the other hand, the logistic regression with pairwise interactions includes all main effects
(just like ours), in addition to $58$ interactions, $5$ of which were in common with our
model, namely the first five listed above. This gives $1+22+58=81$ parameters, i.e.
more than twice as many as our model. Hence, in this case, we obtain a much simpler
model with at least as good performance.

\section{Concluding remarks}\label{sec:concluding_remarks}
We have proposed a generalisation of the simple logistic regression model that can
account for non-linear main effects and complex interactions in the log odds, yet keeping 
the model based on an inherently interpretable structure. This model is constructed
through a specification on generative form, with given marginal distributions from
the natural exponential family, combined with vine copulas to describe the 
dependence. The model parameters are however estimated based on the
discriminative likelihood function, and dependencies between covariates
in the two classes are only included if they contribute significantly to the
discrimination between the two classes. Further, we propose a scheme for
doing model selection and estimation.

To assess the performance of our model, we have conducted a simulation study,
which indicates that our model overall gives good results, also when the sample size
is not very large compared to the number of covariates and in cases where 
non-linearities and interactions are not that strong. We also compared our model
to alternative extensions of the simple logistic regression model, namely a Naive
Bayes model and a logistic regression model with pairwise interactions. Our model
performed either comparably to or better than the others, especially in cases
with strong non-linearities and complex interactions. We also fitted our model 
to two real data sets, namely the ionosphere and the smoking data. Again, the results from our model
were either similar to or better than the compared models. In particular,
the model fitted to the smoking data was much simpler than the best competitor, i.e.
the logistic regression with pairwise interactions, and yet performed just as well.

In the current version of the model selection and estimation procedure, we make an
assumption of strong hierarchy for interactions of order three and higher. This means
that such interactions are only considered for inclusion in the model if all sub interactions
of lower order are already in the model. The reason for this assumption is computational,
as it strongly reduces the number of interactions to check, and thus the computational burden.
It is however straightforward to allow for a weak or no hierarchy instead. Further, our
model is only composed of interactions between continuous covariates. There is however
no explicit way to incorporate interactions between discrete covariates or between 
continuous and discrete ones, and this is a subject for further work.

\section{Acknowledgements}
This work is funded by The Research Council of Norway centre Big Insight, Project 237718.
\bibliography{article3}
\end{document}